\documentclass[10pt]{article}
\usepackage{amsmath, amssymb, amsfonts}
\usepackage{geometry}
\usepackage{graphicx}
\usepackage{hyperref}
\usepackage{enumitem}
\usepackage{booktabs}
\usepackage{physics}
\usepackage{tabularx}
\usepackage{subcaption}
\usepackage{amsthm}
\usepackage{tikz}
\usetikzlibrary{arrows.meta}
\usetikzlibrary{arrows.meta, calc, patterns.meta}

\DeclareMathSymbol{\ii}{\mathalpha}{letters}{"10}
\DeclareMathSymbol{\jj}{\mathalpha}{letters}{"11}
\newcommand{\ihat}{\mathbf {\hat{\ii}}}
\newcommand{\jhat}{\mathbf {\hat{\jj}}}
\newcommand{\khat}{\mathbf {\hat{k}}}

\newcommand{\Lvec}{\mathbf{L}}
\newcommand{\rvec}{\mathbf{r}}

\newcommand{\uvec}{\mathbf{u}}
\newcommand{\Fvec}{\mathbf{F}}
\newcommand{\Ivec}{\mathbf{I}}
\newcommand{\nvec}{\mathbf{n}}
\newcommand{\evec}{\mathbf{e}}
\newcommand{\omvec}{\boldsymbol{\omega}}

\newcommand{\Lamvec}{\boldsymbol{\Omega}}

\begin{document}

\title{Beyond Vorticity: An Angular Momentum Perspective on Fluid Flow}
\author{Ahmed Farooq}
\date{}
\maketitle
\parindent 0ex


\begin{abstract}
While vorticity is the classical tool for analyzing rotational fluid kinematics, it inherently focuses on local, differential spin. This paper introduces a complementary framework based on the angular momentum density field, $\Lvec = \rvec \times \uvec$, deriving generalized transport equations that explicitly balance macroscopic torque and rotational momentum. This $\Lvec$-perspective offers several distinct theoretical advantages over traditional velocity-vorticity formulations. Specifically, this approach: (i) provides a novel decomposition of the viscous torque into a diffusive component and a local spin-dissipative term; (ii) shows the mechanism by which lift is generated in viscous boundary layers by vorticity acting as a source of angular momentum; it also explains stall (iii) reformulates the hydrodynamic impulse to yield a remarkably clean separation of terms into dilatational, volumetric, and rotational flux components; The $\Lvec$ formalism provides the kinematic closure necessary to unify non-circulatory added mass and circulatory lift within a single, dimensionally consistent budget. (iv) enables the direct calculation of the viscous added mass force, accounting for the inertial resistance of boundary layers and separated wakes; (v) simplifies geophysical fluid dynamics by absorbing the planet's rotation—traditionally treated as an artificial virtual vorticity term—directly into the conserved axial angular momentum $m$, revealing the fundamental physics of global circulation through explicit torque balances; (vi) identifies the rotlet as a fundamental Green's function for the $\Lvec$ transport equation in the Stokes regime; and (vii) demonstrates that both oblique shocks and vortex sheets act as singular sources of $\Lvec$ that turn the macroscopic flow. 
\end{abstract}

\section{Introduction}

The Navier-Stokes equations govern the motion of viscous, incompressible fluids, encapsulating conservation of mass and linear momentum. While the vorticity transport equation,  obtained by taking the curl of the momentum equation, describes rotational dynamics. \\

An explicit formulation for angular momentum is less common. The angular momentum per unit volume, $\Lvec$ is defined as:

\begin{equation}
\Lvec = \rvec \times \uvec
\end{equation} 

As can be seen, $\Lvec = \rvec \times \uvec$ is the cross product of the position vector $\rvec$ with the velocity field $\uvec$. The dimension of $\rho \Lvec$ is $[M][L]^{-1} [T]^{-1}$, which coincides with the dimension of angular momentum per unit volume, consistent with its role in capturing rotational dynamics in fluid flows.\\  

The governing equation for $\Lvec$ can be derived by taking a cross product of the Navier Stokes with $\rvec$. This resulting equation, as we will see, highlights the interplay between pressure torques, viscous diffusion, and vorticity, offering a complementary perspective to classical vorticity-based analyses.\\

While $\Lvec$ is widely used in undergraduate textbooks via the Reynolds transport theorem for the analysis of turbomachinery and other flows with a  rotary configuration its reference in other contexts is rather sparse and mostly as a global invariant of motion. For example Wu et al.\footnote{See Eq. 3.7} \cite{wu2007vorticity} have given a relation for  the global angular momentum in terms of vorticity.   Similarly Saffman \cite{Saffman1992} in \S 3.5 gives the global integral for $\Lvec$ and also relates it to vorticity.  Batchelor \cite{Batchelor2000}, while not directly touching upon $\Lvec$ gives the definition of impulse and moments of vorticity in \S 7.2. Truesdell \cite{truesdell2018kinematics} uses $\Lvec$ in his analyis of mean fields of vorticity\footnote{see Chapter 6, \S 64}.\\

In this article we will examine the differential form of the transport equation for $\Lvec$ and study its features and properties.  We will find that it offers fresh insights and perspectives into old problems.  The plan of the article is as follows:  In \S 2 we derive the $\Lvec$ transport equation which gives a unique insight into viscous action and also shows that in the boundary layer limit (in \S 3), it offers an easy interpretation of how lift is generated by turning the flow and also why lift craters in the regime of flow separation. In \S 4 we define various properties of the $\Lvec$ framework and show among other things, how it can be used to calculate circulation. We look at lift generation in inviscid flows in \S 5.  In \S 6 we formulate hydrodynamic impulse in term of $\Lvec$ and find a way to define viscous added mass. In \S 7 we look at geophysical flows and show that in this framework we can get rid of the virtual terrestrial vorticity. In \S 8 we generalize the $\Lvec$ framework to compressible flows.  Finally in \S 9 we show that when applied to Stokes flows, the $\Lvec$ framework shows the ``rotlet" to be the Greens function of the $\Lvec$ transport equation just as the Stokeslet is the Greens function of the $\uvec$ transport equations.  \\

\section{The Governing Equation for $\Lvec$}

Consider the incompressible Navier-Stokes equations for a fluid with constant density $\rho$ and dynamic viscosity $\mu$:

\begin{equation}
\rho \left( \frac{\partial \uvec}{\partial t} + (\uvec \cdot \nabla) \uvec \right) = -\nabla p + \mu \nabla^2 \uvec; \quad \nabla \cdot \uvec = 0
\end{equation}

where $\uvec = (u, v, w)$ is the velocity field and $p$ is the pressure. Define the angular momentum per unit volume as $\Lvec = \rvec \times \uvec$, where $\rvec = (x, y, z)$ is the position vector relative to a fixed origin. It may be noted that the scalar components of the Angular momentum are given as $\Lvec = (L_x, L_y, L_z)$.\\ 

Taking the cross product of the Navier-Stokes momentum equation with $\rvec$, we obtain:

\begin{equation}
\rvec \times \rho \left( \frac{\partial \uvec}{\partial t} + (\uvec \cdot \nabla) \uvec \right) = -\rvec \times \nabla p + \mu \rvec \times \nabla^2 \uvec
\label{eqn:angmom-1}
\end{equation}

Consider the first term $\rvec \times \frac{\partial \uvec}{\partial t}$. Since $\rvec$ is time-independent in a fixed reference frame:

$$
\rvec \times \frac{\partial \uvec}{\partial t} = \frac{\partial}{\partial t} (\rvec \times \uvec) = \frac{\partial \Lvec}{\partial t}
$$

For the convective term, we evaluate the advective derivative of the cross product\footnote{Applying the product rule yields $(\uvec \cdot \nabla)(\rvec \times \uvec) = [(\uvec \cdot \nabla)\rvec] \times \uvec + \rvec \times [(\uvec \cdot \nabla)\uvec]$. Because the directional derivative of the position vector is simply the velocity itself, $(\uvec \cdot \nabla)\rvec = \uvec$, the first term vanishes ($\uvec \times \uvec = \mathbf{0}$). This leaves the exact relation: $\rvec \times [(\uvec \cdot \nabla) \uvec] = (\uvec \cdot \nabla) (\rvec \times \uvec)$.}. Thus, the convective term becomes:\\

$$
\rho \left( \frac{\partial \Lvec}{\partial t} + (\uvec \cdot \nabla) \Lvec \right) = \rho \frac{D \Lvec}{Dt}
$$

Combining the terms we get the equation:

\begin{equation}
\rho \frac{D \Lvec}{Dt} = -\rvec \times \nabla p + \mu \rvec \times \nabla^2 \uvec
\label{eqn:angmom-full-1}
\end{equation}

The above Eq. \ref{eqn:angmom-full-1} shows that quite remarkably, $\Lvec$ also evolves according to a transport equation in the usual form.  The viscous term, $\rvec \times \nabla^2 \uvec$, can be manipulated further using the vector identity:

\begin{equation}
\nabla^2 \Lvec = \nabla^2 (\rvec \times \uvec) = \rvec \times \nabla^2 \uvec + 2 (\nabla \rvec) \times (\nabla \uvec)
\label{eq:nablaL-identity}
\end{equation}

since $\nabla^2 \rvec = 0$. The term $(\nabla \rvec)_{ij} = \delta_{ij}$, so, $
(\nabla \rvec) \times (\nabla \uvec) = \nabla \times \uvec = \omvec$.  Thus:

\begin{equation}
\rvec \times \nabla^2 \uvec = \nabla^2 \Lvec - 2 \omvec
\label{eq:def-vis-mom}
\end{equation}

This relates $\Lvec$ to vorticity. To see this in another form, let us rearrange Eq. \ref{eq:def-vis-mom} as:

\begin{equation}
\omvec  = \frac{1}{2} \left[ \nabla^2 \Lvec - \rvec \times \nabla^2 \uvec \right]
\label{eqn:vort-def}
\end{equation}

We thus see that vorticity can be expressed as half the difference between the Laplace of the angular momentum and the moment of the Laplace of the velocity field.\\

Since the pressure term cannot be simplified any further, the angular momentum transport equation can be written in the form:

\begin{equation}
\rho \frac{D \Lvec}{Dt} = - \rvec \times \nabla p + \mu (\nabla^2 \Lvec - 2 \omvec)
\label{eqn:angmom-full}
\end{equation}

This is the result we have been seeking. As has been noted, the equation has the structure of a transport equation.  Additionally it may be seen that the equation represents the balance of material derivative of the angular momentum and the torques exerted by the pressure and viscous forces.  The material derivative as usual is made of a temporal change and a convective term.\\

This equation has a very close correspondence with the vorticity transport equation which has been reproduced below:

\begin{equation}
\frac{D \omvec}{Dt} = (\omvec \cdot \nabla) \uvec + \nu \nabla^2 \omvec
\end{equation}

As noted previously, vorticity is local concept and represents the ``spin" of the material fluid particle, where as $\Lvec$ represents the angular momentum which has been defined with respect to an origin.  Hence $\Lvec$ is by definition non-local and provides a global measure of the rotational properties of the flow.  \\

Returning to the Laplacian of $\Lvec$ we had the Eq. \ref{eq:def-vis-mom}.  For inviscid irrotational flow, $\omvec = 0$, and viscous diffusion, $\nabla^2 \uvec = 0$, therefore we are left with:

\begin{equation}
\nabla^2 \Lvec =0
\end{equation}

Thus $\Lvec$ field in harmonic for inviscid, irrotational flow.  However for inviscid ($\nabla^2 \uvec =0$) non-irrotational flow, we get:

\begin{equation}
\nabla^2 \Lvec = 2\omvec
\label{eq:def-nabla-L}
\end{equation}

This equation can be solved using Greens functions or other suitable techniques.\\

\subsection{A New Decomposition of the Viscous Torque}

One striking consequence of the angular momentum formalism via Eq.~\ref{eqn:angmom-full} is that it unveils a natural decomposition of the viscous torque into two physically distinct mechanisms. In the Navier-Stokes equation, the influence of viscosity is captured by a single term representing the diffusion viscous force $\mu\nabla^2\uvec$, representing the diffusion of linear momentum. In the vorticity equation, it is the term $\nu\nabla^2\omvec$, representing the diffusion of vorticity. In both cases, the complex physical processes of viscous action are conflated into a single mathematical operator.\\

The $\Lvec$ field transport equation gives two components as: 

\begin{equation}
    \boldsymbol{\tau}_{\text{viscous}} = \mu(\rvec \times \nabla^2\uvec) = \underbrace{\mu\nabla^2 \Lvec}_{\substack{\text{Diffusive} \\ \text{Torque}}} - \underbrace{2\mu \omvec}_{\substack{\text{Local Spin} \\ \text{Torque}}}
    \label{eq:torque_decomposition}
\end{equation}

The first component is the Diffusive Torque, $\mu\nabla^2 \Lvec$. This term is a true spatial diffusion term, mathematically analogous to the diffusion of heat in a solid. It is a non-local effect that depends on the spatial structure of the entire $\Lvec$ field. It does not destroy angular momentum but rather redistributes it through the fluid, causing it to spread from regions of high ``curvature" in the $\Lvec$ field to regions of low ``curvature".\\

The second component is the Local Spin Torque, $-2\mu \omvec$, which is the contribution from the ``spin" of the fluid parcel. It is a purely local term, depending only on the vorticity $\omvec$ at a single point. It acts as a local dissipative brake on the rotation itself, representing a direct frictional drag that removes angular momentum from a fluid parcel in direct proportion to its instantaneous spin rate.\\

\subsection{Behavior of the Viscous Torque in the Limit $Re \to \infty$}

The decomposition of the moments of the viscous torque as shown above leads to different behavior of the two components in the limit of large $Re$ when boundary layers form:\\

The diffusive torque term ($\mu\nabla^2\Lvec$) contains second-order spatial derivatives of the velocity field (since $\Lvec = \rvec \times \uvec$) and therefore, leads to a singular perturbation and ``angular momentum boundary layer" near the surface, where $\Lvec$ is forced to have steep gradients to satisfy no-slip conditions.\\
    
The Local Spin Torque term, ($-2\mu\omvec$) contains only first-order spatial derivatives of velocity  and hence can be expanded as regular perturbation in $Re$. This term represents a distributed, background damping that acts throughout the entire fluid domain, not just in concentrated layers.\\

This separation of the total viscous torque into a singular, boundary-layer-forming part and a regular, distributed-damping part is a unique feature of the angular momentum formalism. It provides a more nuanced and physically accurate description of viscous action than is available from the monolithic viscous terms in the standard momentum or vorticity equations. 

\subsection{Convergence of $\Lvec$ integrals in Unbounded Domains}

An important consideration in the study of angular momentum in unbounded domains is the convergence of the global integral $\int_V \Lvec \, dV$. In three dimensions ($d=3$), the velocity field $\uvec$ of a localized disturbance typically behaves as a dipole, decaying as $O(r^{-3})$. Consequently, the angular momentum density $\Lvec = \rvec \times \uvec$ decays as $O(r^{-2})$. As noted by Saffman \cite{Saffman1992} and others, the volume integral of such a field is not absolutely convergent and may exhibit shape-dependency as the boundary $S_{\infty}$ tends to infinity.\\

Caution therefore is warranted regarding the convergence of global integrals involving the $\Lvec$-field, and as we will see in subsequent sections that for various applications, it is advantageous to utilize the divergence $\nabla \cdot \Lvec$ or the curl $\nabla \times \Lvec$, etc. to resolve non-local flow features. In many such instances, we leverage the gradient or divergence theorems to map potentially divergent global volume integrals onto well-defined body surface integrals, thereby regularizing the kinematic fields and ensuring a finite result. We will explicitly address these convergence considerations and the utility of these vector operators as the specific physical scenarios arise throughout the remainder of this work.

\section{Boundary Layers and Lift Generation}

At high $Re$ the no-slip surface condition at the wall creates an ``angular momentum'' boundary layer. It implies in the usual manner that high gradients exist in a thin region near a solid surface where viscous effects remain important and the field variables change rapidly, even as the global viscosity $\mu$ becomes very small. For general references on Boundary layers, Rosenhead \cite{Rosenhead1963} and Schlichting \cite{schlichting2016boundary} may be consulted.\\

Considering a steady, 2D, high-Reynolds number flow over a flat plate at $y=0$, with a freestream velocity $U$. The full, steady-state transport equation for the axial component of specific angular momentum, $L_z = xv - yu$, (Eq. \ref{eqn:angmom-full}) is:

\begin{equation}
u\frac{\partial L_z}{\partial x} + v\frac{\partial L_z}{\partial y} = -\frac{1}{\rho}\left(x\frac{\partial p}{\partial y} - y\frac{\partial p}{\partial x}\right) + \nu\left(\frac{\partial^2 L_z}{\partial x^2} + \frac{\partial^2 L_z}{\partial y^2} - 2\omega_z\right)
\label{eq:full_Lz_transport}
\end{equation}

Applying the standard Prandtl boundary layer scaling assumptions, where the boundary layer thickness $\delta$ is much smaller than the characteristic length $L_0$: $x \sim L_0$, \quad $y \sim \delta$,    $u \sim U$, \quad $v \sim U(\delta/L_0)$,  $\partial/\partial x \sim 1/L_0$, $\quad \partial/\partial y \sim 1/\delta$,   $\nu \sim \delta^2 U / L_0$.  Within the boundary layer $L_z \sim U\delta$ and $\omega_z \sim U/\delta$. Under this scaling, we arrive at a new, simplified equation:

\begin{equation}
u\frac{\partial L_z}{\partial x} + v\frac{\partial L_z}{\partial y} = \nu\left(\frac{\partial^2 L_z}{\partial y^2} - 2\omega_z\right)
\label{eq:AMBL}
\end{equation}

An important feature of Eq. \ref{eq:AMBL} is that the vorticity ($-2\nu\omega_z$) acts like a source term for the $\Lvec$ field.  This provides a neat explanation for how lift is generated which is not intuitively obvious in the traditional velocity-vorticity framework:  Let us consider flow over an airfoil, then as is well known,  the surface of the foil acts as a source of vorticity.  But in the velocity-vorticity framework, there is no way to see how this vorticity generates lift.  But Eq. \ref{eq:AMBL} provides a simple explanation:  since the vorticity is a source of angular momentum, this generates a net turning of the flow and lift is the reaction on the airfoil.  We consider this example in detail below.\\

\subsection{Vorticity Generation and Asymmetry at an Angle of Attack}

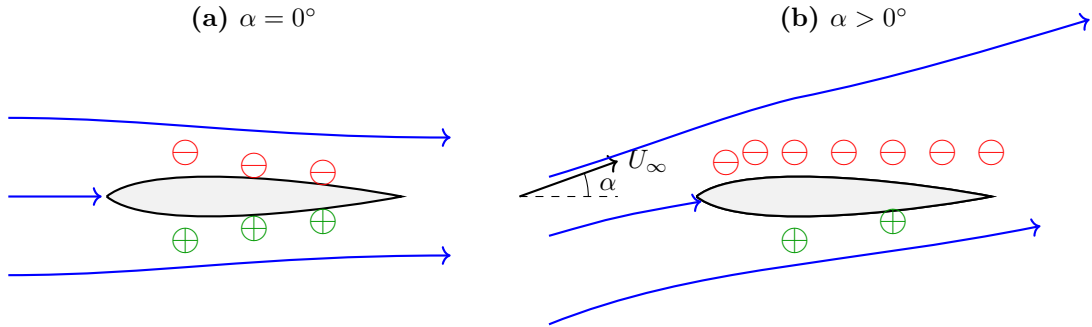
\begin{figure}[h]
\centering
\begin{tikzpicture}[scale=1.3]
    \begin{scope}
        \node at (1.5, 1.8) {\textbf{(a) $\alpha = 0^\circ$}};
        
        \draw[thick, fill=gray!10] (0,0) .. controls (0.5, 0.4) and (2.5, 0.1) .. (3,0) .. controls (2.5, -0.1) and (0.5, -0.4) .. (0,0);
        
        \draw[blue, thick, ->] (-1, 0.8) .. controls (0.5, 0.8) and (1.5, 0.6) .. (3.5, 0.6);
        \draw[blue, thick, ->] (-1, -0.8) .. controls (0.5, -0.8) and (1.5, -0.6) .. (3.5, -0.6);
        \draw[blue, thick, ->] (-1, 0.0) -- (-0.05, 0); 

        \node[red, font=\Large] at (0.8, 0.45) {$\ominus$};
        \node[red, font=\Large] at (1.5, 0.32) {$\ominus$};
        \node[red, font=\Large] at (2.2, 0.25) {$\ominus$};
        
        \node[green!60!black, font=\Large] at (0.8, -0.45) {$\oplus$};
        \node[green!60!black, font=\Large] at (1.5, -0.32) {$\oplus$};
        \node[green!60!black, font=\Large] at (2.2, -0.25) {$\oplus$};
    \end{scope}

    \begin{scope}[shift={(6, 0)}]
    \node at (1.5, 1.8) {\textbf{(b) $\alpha > 0^\circ$}};    
    
    \draw[dashed] (-1.8, 0) -- (-0.8, 0);
    \draw[thick, ->] (-1.8, 0) -- (-0.8, 0.36) node[right] {$U_\infty$};
    \draw (-1.1, 0) arc[start angle=0, end angle=20, radius=0.7];
    \node at (-0.9, 0.12) {$\alpha$};

    \draw[thick, fill=gray!10] (0,0) .. controls (0.5, 0.4) and (2.5, 0.1) .. (3,0) .. controls (2.5, -0.1) and (0.5, -0.4) .. (0,0);
    
    \node[red, font=\Large] at (0.3, 0.35) {$\ominus$};
    \node[red, font=\Large] at (0.6, 0.45) {$\ominus$};
    
    \node[red, font=\Large] at (1.0, 0.45) {$\ominus$};
    \node[red, font=\Large] at (1.5, 0.45) {$\ominus$};
    \node[red, font=\Large] at (2.0, 0.45) {$\ominus$};
    \node[red, font=\Large] at (2.5, 0.45) {$\ominus$};
    \node[red, font=\Large] at (3.0, 0.45) {$\ominus$};


    \draw[thick, fill=gray!10] (0,0) .. controls (0.5, 0.4) and (2.5, 0.1) .. (3,0) .. controls (2.5, -0.1) and (0.5, -0.4) .. (0,0);
    
    \node[green!60!black, font=\Large] at (1.0, -0.45) {$\oplus$};
    \node[green!60!black, font=\Large] at (2.0, -0.25) {$\oplus$};
    
    \draw[blue, thick, ->] (-1.5, 0.2) .. controls (-0.5, 0.5) and (0.2, 0.8) .. (1.0, 1.0) .. controls (2.0, 1.2) and (3.0, 1.5) .. (4.0, 1.8);
    \draw[blue, thick, ->] (-1.5, -1.3) .. controls (0, -0.7) and (1.5, -0.7) .. (3.5, -0.3);
    \draw[blue, thick, ->] (-1.5, -0.4) .. controls (-0.8, -0.2) .. (0.05, -0.05);

    \end{scope}
\end{tikzpicture}
\caption{Schematic of boundary layer vorticity generation on a symmetric airfoil. Red minus signs ($-$) denote clockwise (negative) vorticity; green plus signs ($+$) denote counter-clockwise (positive) vorticity. At $\alpha > 0^\circ$, the upper surface flow rapidly accelerates, creating a severe velocity gradient that generates a dominant sheet of negative vorticity.}
\label{fig:vorticity_aoa}
\end{figure}

When we have a symmetric airfoil at zero angle of attack ($\alpha =0$), it leads to symmetric flow (see Fig. \ref{fig:vorticity_aoa}(a)). Since vorticity acts as a source of angular momentum, we must consider the vorticity generation mechanism on the upper and lower surfaces of the airfoil. The spanwise vorticity ($\omega_z$) within a thin attached boundary layer is driven almost entirely by the wall-normal velocity gradient:

\begin{equation}
\omega_z \approx -\frac{\partial u}{\partial y} =- \frac{\Delta u}{\Delta y}
\label{eq:vort-approx}
\end{equation}

On the upper surface, the fluid velocity must transition from the no-slip condition ($u=0$) at the wall to a highly accelerated external velocity ($U_\infty$) over a very short distance. This extremely steep velocity gradient generates a correspondingly intense sheet of negative (clockwise) vorticity (since both $\Delta u$ and $\Delta y$ in Eq.~\ref{eq:vort-approx} are positive).\\

However on the lower surface of the airfoil, $\omega_z \approx - \frac{\Delta u}{\Delta y}$, but $\Delta y$ is negative so it results in positive vorticity generation.  The distribution of the vorticity is perfectly antisymmetric around the airfoil with the result that the net angular momentum imparted to the flow (which as per Eq.~\ref{eq:AMBL} is sourced by the vorticity), is therefore zero. Since the flow doesn't turn, there is no reaction on the airfoil and hence there is no lift generation.

When an airfoil is pitched to a positive angle of attack (see Fig. \ref{fig:vorticity_aoa}(b)), the oncoming fluid stagnates slightly under the leading edge. Fluid destined for the upper surface must negotiate a much tighter, highly convex geometric path. This extreme curvature induces a severe local pressure drop, causing the upper surface flow to accelerate significantly past the free-stream velocity ($U_\infty$). Conversely, the lower surface presents a shallower geometric profile, and the flow coasts along it at a slightly decelerated speed.\\

On the lower surface, the transition from the wall to a slower external velocity yields a much shallower gradient, producing a significantly weaker sheet of positive (counter-clockwise) vorticity.\\

This imbalance in the asymmetric distribution of vorticity creates a net negative source of vorticity within the boundary layer which acts as a ``source" of angular momentum as given by Eq.~\ref{eq:AMBL} turning the flow downward.  As a reaction to the to downward turning of the flow we see a positive lift generation. \\

The airfoil is, in effect, functioning as a fluidic torque generator. By Newton's Third Law, the sustained ejection of this rotational debt requires an equal and opposite reaction force on the solid body. It is this continuous, asymmetric advection of the moment of momentum that we physically measure as aerodynamic lift.\\

We make these ideas exact in the next section.

\subsection{The Source of Rotational Debt: Linking Vorticity to Angular Momentum Flux}

To rigorously bridge classical circulation theory with the $\Lvec$-framework, we examine the steady-state transport of angular momentum within the boundary layer. Starting with the angular momentum transport boundary layer equation (Eq.~\ref{eq:AMBL}):

\[
u\frac{\partial L_z}{\partial x} + v\frac{\partial L_z}{\partial y} = \nu\left(\frac{\partial^2 L_z}{\partial y^2} - 2\omega_z\right)
\]

To determine the global aerodynamic footprint of the wing, we integrate this transport equation over a control volume $V$ encompassing the entire boundary layer and the immediate near-wake. \\

First, we rewrite the advective left-hand side in conservative form. Applying the continuity equation ($\nabla \cdot \uvec = 0$), the left-hand side becomes the exact divergence of the angular momentum flux: $\nabla \cdot (\uvec L_z)$. The transport equation over the volume $V$ is therefore:

\begin{equation}
\iint_V \nabla \cdot (\uvec L_z) \, dV = \iint_V \nabla \cdot (\nu \nabla L_z) \, dV - \iint_V 2\nu \omega_z \, dV
\end{equation}

Next, we apply the Divergence Theorem to convert the volume integrals of the divergences into surface fluxes across the boundary $S$ (which consists of the solid airfoil surface $S_{body}$ and the outer fluid boundary $S_{out}$):

\begin{equation}
\oint_S L_z (\uvec \cdot \nvec) \, dA = \oint_S \nu \nabla L_z \cdot \nvec \, dA - 2\nu \iint_V \omega_z \, dV
\label{eq:rotationl-debt}
\end{equation}

This integrated form reveals a striking physical balance. The term on the left-hand side is exactly the net convective flux of angular momentum out of the control volume—this is the \textit{rotational debt}  continuously released into the fluid to sustain lift.\\

The volume integral of the vorticity ($\iint_V \omega_z dV$) is, by definition, the total bound circulation of the airfoil ($\Gamma_{v,bound}$\footnote{The $v$ is a reminder that this is bound vorticity for fully viscous flow.}). Therefore, the equation simplifies to:

\begin{equation}
\text{Rotational Debt Flux} = -2\nu \Gamma_{v,bound} + \text{Viscous Diffusion of } L_z
\label{eq:stall}
\end{equation}

The importance of the above equation cannot be overstated because it clearly links the angular momentum flux to the vorticity integral. It should be noted that $\Gamma_{v,bound}$ is defined for viscous flows as a volume integral of vorticity.   The equation demonstrates that the volume integral of the boundary layer vorticity—less the angular momentum that strictly diffuses away through viscous action at the boundaries—yields the exact rotational debt released into the fluid. 

\subsection{The Lift Integral}

To  connect the rotational debt framework to the generation of lift, we must evaluate the momentum balance in the vertical direction over the fluid domain. Consider a 2D control volume $V$ bounded internally by the airfoil surface ($S_{\text{airfoil}}$) and externally by an infinite boundary ($S_\infty$) as $x, y \to \pm\infty$. The lift, $F_L$, requires calculating the vertical momentum flux across the downstream boundary:

\begin{equation}
F_L = \int_{-\infty}^{\infty} \rho u v \, dy
\label{eq:cont-vol-mom}
\end{equation}

However, because the fluid is incompressible, the localized vorticity of the airfoil induces a global velocity field. The vertical induced velocity decays asymptotically ($v \sim 1/r$), extending infinitely deep into the irrotational regions. Consequently, the momentum integral given by Eq.~\ref{eq:cont-vol-mom} is conditionally convergent, rendering the control volume approach unreliable.\\

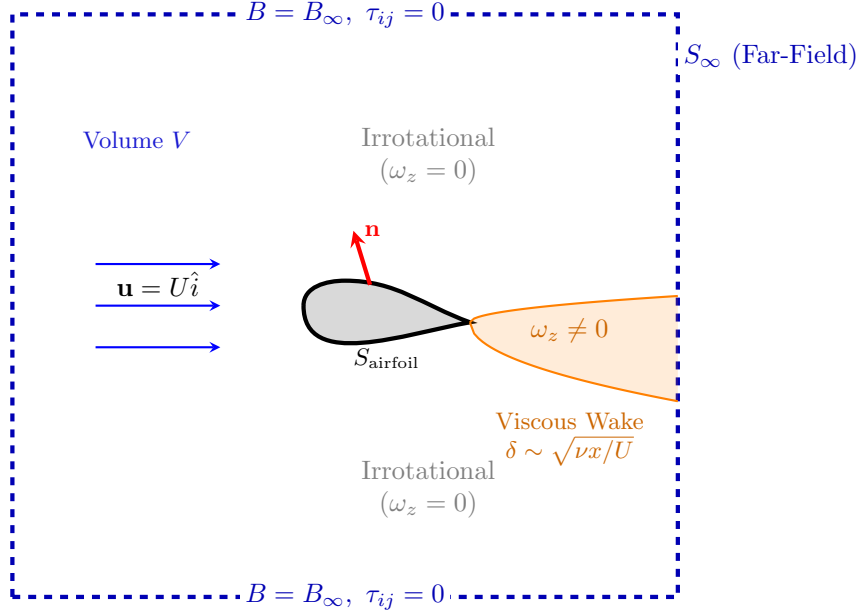
\begin{figure}[htbp]
\centering
\begin{tikzpicture}[>=stealth, scale=1.1]
    
    \draw[dashed, ultra thick, blue!70!black] (-3.5, -3.5) rectangle (4.5, 3.5);
    \node[blue!70!black, right, fill=white, inner sep=2pt] at (4.5, 3) {$S_\infty$ (Far-Field)};
    
    \node[blue!70!black, fill=white, inner sep=2pt] at (0.5, 3.5) {$B = B_\infty, \ \tau_{ij} = 0$};
    \node[blue!70!black, fill=white, inner sep=2pt] at (0.5, -3.5) {$B = B_\infty, \ \tau_{ij} = 0$};
    
    \node[font=\small, blue!80!black, opacity=0.9] at (-2, 2) {Volume $V$};
    
    \filldraw[fill=gray!30, draw=black, ultra thick] (0,0) 
        to[out=90, in=180] (0.5, 0.3) 
        to[out=0, in=160] (2, -0.2) 
        to[out=190, in=270] (0,0);
    \node[black, font=\small, below] at (1, -0.4) {$S_{\text{airfoil}}$};
    
    \draw[->, red, ultra thick] (0.8, 0.25) -- (0.6, 0.9) node[right, font=\small] {$\mathbf{n}$};
        
    \draw[->, blue, thick] (-2.5, 0.5) -- (-1, 0.5);
    \draw[->, blue, thick] (-2.5, 0) -- (-1, 0) node[midway, above, text=black] {$\mathbf{u} = U\hat{i}$};
    \draw[->, blue, thick] (-2.5, -0.5) -- (-1, -0.5);
    
    \draw[thick, orange, domain=2:4.5, samples=50] plot (\x, {0.2*sqrt(\x-2) - 0.2});
    \draw[thick, orange, domain=2:4.5, samples=50] plot (\x, {-0.6*sqrt(\x-2) - 0.2});
    
    \fill[orange, opacity=0.15, domain=2:4.5, variable=\x] 
        (2,-0.2) -- plot (\x, {0.2*sqrt(\x-2) - 0.2}) -- (4.5, 0.116) -- 
        (4.5, -1.148) -- plot[domain=4.5:2] (\x, {-0.6*sqrt(\x-2) - 0.2}) -- cycle;
        
    \node[orange!80!black] at (3.2, -0.3) {$\omega_z \neq 0$};
    \node[align=center, text=orange!80!black, font=\small] at (3.2, -1.6) {Viscous Wake\\ $\delta \sim \sqrt{\nu x/U}$};
    
    \node[gray, align=center] at (1.5, 1.8) {Irrotational\\ ($\omega_z = 0$)};
    \node[gray, align=center] at (1.5, -2.2) {Irrotational\\ ($\omega_z = 0$)};

\end{tikzpicture}
\caption{Control volume representation of the Lamb vector integration. The fluid volume $V$ is bounded internally by the solid airfoil surface ($S_{\text{airfoil}}$) and externally by an infinite far-field boundary ($S_\infty$). Because the irrotational far-field features a uniform Bernoulli pressure ($B = B_\infty$) and zero viscous stress, the surface integral on $S_\infty$ identically vanishes. The application of Gauss's theorem strictly maps the localized volume integral of the wake ($\int \rho u \omega_z \, dV$) directly to the lift force generated at the physical boundary $S_{\text{airfoil}}$.}
\label{fig:volume_integral_gauss}
\end{figure}

To circumvent these convergence problems  and formally connect the rotational debt framework to the macroscopic generation of lift, we build upon the exact aerodynamic force formalisms pioneered by Wu \cite{wu1981theory} and subsequently generalized by Noca \cite{noca1999comparison}. By adopting their fundamental methodology of mapping classical boundary momentum fluxes to localized Lamb vector volume integrals (see Fig.~\ref{fig:volume_integral_gauss}), we can evaluate the lift force balance over the fluid domain while bypassing the conditionally convergent velocity tails mentioned above. We start by recasting the convective acceleration in the steady, incompressible Navier-Stokes equations via the identity:

\begin{equation}
(\mathbf{u} \cdot \nabla)\mathbf{u} = \boldsymbol{\omega} \times \mathbf{u} + \frac{1}{2}\nabla|\mathbf{u}|^2
\end{equation}
 
the static pressure and kinetic energy gradient combine into the total Bernoulli pressure, $B = p + \frac{1}{2}\rho|\mathbf{u}|^2$. The momentum balance becomes:

\begin{equation}
\rho (\boldsymbol{\omega} \times \mathbf{u}) = -\nabla B + \mu \nabla^2 \mathbf{u}
\end{equation}

Taking the transverse ($y$) component of this equation—where the cross product simplifies to $\omega_z u$—and integrating over the entire infinite fluid volume $V$ yields:

\begin{equation}
\int_V \rho u \omega_z \, dV = -\int_V \frac{\partial B}{\partial y} \, dV + \mu \int_V \nabla^2 v \, dV
\label{eq:vol-1}
\end{equation}

To show that the right-hand side of this volume integral equates exactly to the aerodynamic lift ($F_L$), we apply Gauss's Divergence Theorem. The volume integral of the spatial derivatives transforms into a surface integral evaluated over the boundaries bounding the fluid: the infinite far-field ($S_\infty$) and the solid airfoil surface ($S_{\text{airfoil}}$). Defining $\mathbf{n}$ as the normal vector pointing outward from the airfoil into the fluid, the right-hand side becomes:

\begin{equation}
\text{RHS} = \oint_{S_\infty} \left( \dots \right) dS - \oint_{S_{\text{airfoil}}} \left( -B n_y + \mu (\nabla v \cdot \mathbf{n}) \right) \, dS
\end{equation}

As established, the integration over the irrotational far-field boundary ($S_\infty$) evaluates identically to zero. The analytical closure occurs on the solid boundary $S_{\text{airfoil}}$. By enforcing the viscous no-slip condition, the local fluid velocity is strictly zero ($|\mathbf{u}| = 0$). Consequently, the dynamic pressure component vanishes, and the total Bernoulli pressure collapses exactly to the local static pressure ($B \to p$). \\

The surviving surface integral evaluates the static pressure and the $y$-component of the viscous velocity gradient across the geometry of the wing:

\begin{equation}
\text{RHS} = \oint_{S_{\text{airfoil}}} \left( p n_y - \mu (\nabla v \cdot \mathbf{n}) \right) \, dS \equiv F_L
\end{equation}

By Newton's formulation, the integration of normal static pressure and tangential viscous shear over the solid boundary constitutes the exact definition of the aerodynamic force. Thus, the right-hand side formally equates to $F_L$, proving unequivocally that the volume integral of the Lamb vector ($\int_V \rho u \omega_z \, dV$) is identically equal to the  lift.

\begin{equation}
F_L = \int_V \rho u \omega_z \, dV
\label{eq:vortex-force}
\end{equation}

This formulation shows that because the vorticity $\omega_z$ is identically zero everywhere outside the boundary layer and viscous wake, the mathematically tricky infinite volume limits strictly collapse to a finite envelope ($y \in [-\delta, \delta]$, $\delta \sim \nu x/U$ is the diffusive length scale, which makes the vorticity distribution compact). \\

To map the result given by Eq.~\ref{eq:vortex-force} to a form useful in Eq.~\ref{eq:rotationl-debt} we decompose the streamwise velocity field inside the wake. Let the local velocity $u$ be expressed as the uniform freestream $U$ perturbed by a localized viscous velocity deficit $u'$ ($u = U - u'$). Substituting this into the exact Lamb integral yields:

\begin{equation}
F_L = \rho U \int_V \omega_z \, dV - \rho \int_V u' \omega_z \, dV
\label{eq:modified-KJ}
\end{equation}

The above equation illuminates the mechanics of the rotational debt framework with the first term representing the dominant, linear contribution. Because $U$ is spatially uniform, it factors out, leaving the volume integral of the raw vorticity field, which is exactly the total circulation ($\int_V \omega_z \, dV \equiv \Gamma_{v,bound}$). This leading-order term exactly recovers the classical Kutta-Joukowski theorem ($F_L = \rho U \Gamma_{v,bound}$) directly from the fully viscous equations, without invoking inviscid assumptions or artificial contour selections.  We thus have:

\begin{equation}
F_L = \rho U \Gamma_{v,bound} - \rho \int_V u' \omega_z \, dV
\label{eq:modified-KJ-1}
\end{equation}

The second term represents a non-linear viscous correction, accounting for the reality that vorticity is transported downstream at a local convective velocity slightly slower than the freestream. \\

So now by Eq.~\ref{eq:modified-KJ-1} and Eq.~\ref{eq:rotationl-debt} we have related the rotational debt $\nabla \cdot (\uvec L_z)$ to the lift force $F_L$, making exact our intuition about how vorticity turns the flow and generates lift.\\

For streamlined airfoils with thin boundary layers and a narrow trailing wake, the velocity deficit is a small perturbation relative to the freestream ($u' \ll U$) and the second term in Eq.~\ref{eq:modified-KJ-1} can be neglected. However this may not be possible for bluff bodies or separated flows. 

\subsection{Flow Separation and the Collapse of Lift}

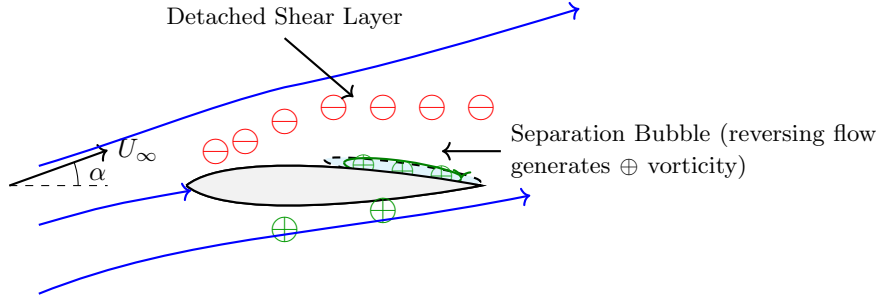
\begin{figure}[h]
\centering
\begin{tikzpicture}[scale=1.3]
    
    \draw[dashed] (-1.8, 0) -- (-0.8, 0);
    \draw[thick, ->] (-1.8, 0) -- (-0.8, 0.36) node[right] {$U_\infty$};
    \draw (-1.1, 0) arc[start angle=0, end angle=20, radius=0.7];
    \node at (-0.9, 0.12) {$\alpha$};

    \draw[thick, fill=gray!10] (0,0) .. controls (0.5, 0.4) and (2.5, 0.1) .. (3,0) .. controls (2.5, -0.1) and (0.5, -0.4) .. (0,0);
    
    \node[red, font=\Large] at (0.3, 0.35) {$\ominus$};
    \node[red, font=\Large] at (0.6, 0.45) {$\ominus$};
    
    \node[red, font=\Large] at (1.0, 0.65) {$\ominus$};
    \node[red, font=\Large] at (1.5, 0.8) {$\ominus$};
    \node[red, font=\Large] at (2.0, 0.8) {$\ominus$};
    \node[red, font=\Large] at (2.5, 0.8) {$\ominus$};
    \node[red, font=\Large] at (3.0, 0.8) {$\ominus$};

    
    \begin{scope}[shift={(0.4, -0.20)}, rotate around={-8:(1.8, 0.35)}]
        \draw[thick, dashed, fill=cyan!10] (1.8, 0.35) ellipse (0.8 and 0.08);
        
        \draw[->, thick, green!50!black] (1.2, 0.35) arc[start angle=180, end angle=0, x radius=0.6, y radius=0.1];
        \draw[thick, green!50!black] (2.4, 0.35) arc[start angle=0, end angle=-180, x radius=0.6, y radius=0.1];
        
        \node[green!60!black, font=\large] at (1.4, 0.35) {$\oplus$};
        \node[green!60!black, font=\large] at (1.8, 0.35) {$\oplus$};
        \node[green!60!black, font=\large] at (2.2, 0.35) {$\oplus$};
    \end{scope}
    
    \draw[thick, fill=gray!10] (0,0) .. controls (0.5, 0.4) and (2.5, 0.1) .. (3,0) .. controls (2.5, -0.1) and (0.5, -0.4) .. (0,0);
    
    \node[green!60!black, font=\Large] at (1.0, -0.45) {$\oplus$};
    \node[green!60!black, font=\Large] at (2.0, -0.25) {$\oplus$};
    
    \draw[blue, thick, ->] (-1.5, 0.2) .. controls (-0.5, 0.5) and (0.2, 0.8) .. (1.0, 1.0) .. controls (2.0, 1.2) and (3.0, 1.5) .. (4.0, 1.8);
    \draw[blue, thick, ->] (-1.5, -1.1) .. controls (0, -0.5) and (1.5, -0.5) .. (3.5, -0.1);
    \draw[blue, thick, ->] (-1.5, -0.4) .. controls (-0.8, -0.2) .. (0.05, -0.05);

    \draw[<-, thick] (1.7, 0.9) -- (1.0, 1.5) node[above] {\small Detached Shear Layer};
    \draw[<-, thick] (2.6, 0.35) -- (3.2, 0.35) node[right, text width=5.00cm, align=left] {\small Separation Bubble  (reversing flow generates $\oplus$ vorticity)};
\end{tikzpicture}
\caption{Schematic of boundary layer separation at high angles of attack, shown in the body-axis frame. The incoming flow ($U_\infty$) approaches at angle $\alpha$. The severe adverse pressure gradient over the aft section causes the negative vorticity ($-$) shear layer to detach. Beneath it, a distinct separation bubble forms. The recirculating (clockwise) flow inside this bubble acts as a generator of positive vorticity ($+$), which cannibalizes the net circulation budget and destroys aerodynamic lift.}
\label{fig:stall}
\end{figure}

The asymmetric generation of vorticity—and the resulting lift—relies entirely on the flow remaining attached to the upper surface as we have seen previously. As the angle of attack increases, the fluid accelerating over the leading edge must subsequently decelerate as it progresses toward the trailing edge, moving against a severe adverse pressure gradient. At a critical angle, the viscous boundary layer expends its kinetic energy and detaches from the surface, initiating aerodynamic stall (Fig. \ref{fig:stall}).  We now understand how stall leads to the fall in lift using Eq.~\ref{eq:stall}.\\

When separation occurs, the velocity gradient on the upper solid surface is weakened. The sheet of negative vorticity detaches and is carried away in a free shear layer. Beneath this detached layer, a separation bubble forms. Within this wake, the fluid recirculates, meaning the local velocity near the wall reverses direction and flows upstream against the free-stream. \\

Since the fluid velocity is negative near the wall and positive further out in the shear layer, the local velocity gradient ($\partial u/\partial y$) becomes strongly positive. This reversed flow generates a positive (counter-clockwise) vorticity directly on the upper surface. This is depicted in Fig. \ref{fig:stall}.\\

From our perspective, this newly generated positive vorticity effectively cannibalizes the remaining negative vorticity in the boundary layer budget on the upper surface and reduces the ``net" contribution of the vorticity source term $- 2\nu \iint_V \omega_z \, dV$ which is essential for lift generation. Thus the imbalance that previously defined the lifting airfoil is neutralized, reducing the net bound circulation ($\Gamma$) and driving it toward zero.\\

This is a clear and direct explanation for the mechanism of how stall causes the the collapse of lift generation.  We will see this again in the next section in the context of vorticity sheets where we consider inviscid flows.

\section{Some Properties of the Angular Momentum Field}

This section examines a few of the useful properties and the concomitant dynamics of the angular momentum field, $\Lvec $, building upon its transport equation derived in the previous section. We explore its divergence and curl and also derive the generalized energy associated with the field.  We also look at the role of Helicity and Lambs vector from this lens.  Further we discuss the crucial role of the origin's choice in its definition.

\subsection{Divergence and Curl of $\Lvec$}

Understanding the spatial distribution of angular momentum is crucial for characterizing rotational fluid flows. Here we derive a fundamental relationship between the divergence of the angular momentum per unit volume and the flow's vorticity distribution. This relationship provides insights into regions where angular momentum effectively ``originates" or ``terminates" within the flow, drawing an intriguing analogy to electrostatics.\\

We begin by computing the divergence  of $\Lvec$:

\begin{equation}
\nabla \cdot \Lvec = \nabla \cdot (\rvec \times \uvec) 
\end{equation}

Applying the vector identity: $\nabla \cdot (\mathbf{A} \times \mathbf{B}) = \mathbf{B} \cdot (\nabla \times \mathbf{A}) - \mathbf{A} \cdot (\nabla \times \mathbf{B}) $ we obtain:

\begin{equation}
\nabla \cdot (\rvec \times \uvec) = \uvec \cdot (\nabla \times \rvec) - \rvec \cdot (\nabla \times \uvec)
\end{equation}

We recall that the curl of the position vector is identically zero ($\nabla \times \rvec = \mathbf{0}$) and that the vorticity vector is defined as the curl of the velocity field ($\omvec = \nabla \times \uvec$). Substituting these into the expression for $\nabla \cdot \Lvec$ leads to the fundamental identity:

\begin{equation}
\nabla \cdot \Lvec = -\rvec \cdot \omvec
\label{eq:div-L}
\end{equation}

Eq. \ref{eq:div-L} demonstrates dependence of  the divergence on the vorticity distribution in any incompressible flow. This identity is universally valid for such flows, independent of specific boundary conditions or external forces. The term $\rvec \cdot \omvec$ quantifies the extent to which vorticity aligns with the position vector.\\

\textbf{Curl of Angular Momentum}\\

The curl of the angular momentum field, $\nabla \times \Lvec = \nabla \times (\rvec \times \uvec)$, is computed using  the general vector identity for the curl of a cross product: $ \nabla \times (\mathbf{A} \times \mathbf{B}) = \mathbf{A} (\nabla \cdot \mathbf{B}) - \mathbf{B} (\nabla \cdot \mathbf{A}) + (\mathbf{B} \cdot \nabla) \mathbf{A} - (\mathbf{A} \cdot \nabla) \mathbf{B} $  and continuity to give:

\begin{equation}
\nabla \times \Lvec = -2 \uvec - (\rvec \cdot \nabla) \uvec
\label{eq:curl-L1}
\end{equation}

Eq. \ref{eq:curl-L1} establishes a fundamental relationship between the curl of the angular momentum density, the local velocity, and the directional derivative of the velocity field along the position vector. This result is valid for all 3D incompressible flows. It highlights that the rotational characteristics of the angular momentum field are governed not only by the velocity itself but also by how the velocity changes with position relative to the origin.\\ 

\subsection{Poisson Equation for $\Lvec$}

We investigate the behavior and utility of the angular momentum, $\Lvec$, within the context of inviscid fluid flows in this section and also sections 4.3 and 4.4. This flow regime leads to significant simplification of the governing equations by setting the viscosity $\mu = 0$. For general references on inviscid flow, Kochin \cite{Kochin1964}, Milne-Thompson \cite{MilneThomson1968}, Katz and Plotkin \cite{KatzPlotkin2001},  Lighthill \cite{Lighthill1986} may be consulted. \\

For inviscid rotational flows the flow is determined by the Poisson equation, Eq. \ref{eq:def-nabla-L}:

\begin{equation}
\nabla^2 \Lvec = 2 \omvec
\end{equation}

We examine the case where $\omvec = \delta^{(2)}(\rvec - \mathbf{r_0})$, giving the Greens function as:

\begin{equation}
G(\rvec,\mathbf{r_0}) = \frac{1}{2 \pi} \frac{1}{|\rvec - \mathbf{r_0}|}
\end{equation}

When the flow is  irrotational or where $\nabla^2 \uvec = 0$, the previously derived equations for $\Lvec$ simplify considerably. Specifically, the Laplacian of $\Lvec$ becomes:

\begin{equation}
\nabla^2 \Lvec = 0 
\end{equation}

Furthermore, the divergence relationship derived earlier simplifies for irrotational flows to:

\begin{equation}
\nabla \cdot \Lvec = 0 
\end{equation}

Next, we develop some of the well known results of inviscid irrotational theory in the $\Lvec$ framework.

\subsection{Influence of Origin Location on Angular Momentum Properties}
The definition of angular momentum, $\Lvec = \rvec \times \uvec$, inherently depends on the choice of the origin for the position vector $\rvec$. This subsection explores how a shift in the chosen origin affects the angular momentum per unit volume and its related differential properties (divergence, curl, and Laplacian). Understanding this dependence is crucial for appropriate interpretation of $\Lvec$ in various flow configurations and for simplifying analytical or computational efforts.\\

Let the original origin be $\mathbf{O} = (0, 0, 0)$, and let the new origin be $\mathbf{O}' = \boldsymbol{\chi} = (\chi_x, \chi_y, \chi_z)$, a constant vector representing the displacement of the new origin from the old. The position vector relative to the new origin, $\rvec'$, is then given by $\rvec' = \rvec - \boldsymbol{\chi}$. Consequently, the angular momentum per unit volume with respect to the new origin, $\Lvec'$, transforms as:

\[ 
\Lvec' = \rvec' \times \uvec = (\rvec - \boldsymbol{\chi}) \times \uvec = \rvec \times \uvec - \boldsymbol{\chi} \times \uvec = \Lvec - \boldsymbol{\chi} \times \uvec 
\]

This fundamental transformation demonstrates that a shift in the chosen origin introduces an additional velocity-dependent term, $-\boldsymbol{\chi} \times \uvec$.\\

The transformations of the angular momentum and its differential properties under an origin shift are summarized in Table \ref{tab:L_transformations}.\\

\begin{table}[h!]
\centering
\caption{Transformations under Origin Shift $\rvec' = \rvec - \boldsymbol{\chi}$}
\label{tab:L_transformations}
\begin{tabular}{l c}
\toprule
\textbf{Quantity} & \textbf{Transformation Rule} \\
\midrule
Angular Momentum ($\Lvec'$) & $\Lvec' = \Lvec - \boldsymbol{\chi} \times \uvec$ \\
\addlinespace[0.5em]
Divergence ($\nabla \cdot \Lvec'$) & $\nabla \cdot \Lvec' = -\rvec' \cdot \omvec$ \\
\addlinespace[0.5em]
Curl ($\nabla \times \Lvec'$) & $\nabla \times \Lvec' = -2 \uvec - (\rvec' \cdot \nabla) \uvec$ \\
\addlinespace[0.5em]
Laplacian ($\nabla^2 \Lvec'$) & $\nabla^2 \Lvec' = \nabla^2 \Lvec - \boldsymbol{\chi} \times \nabla^2 \uvec$ \\
\bottomrule
\end{tabular}
\end{table}

The table illustrates that while the form of the divergence and curl relations remains invariant with respect to the chosen origin (i.e., $\rvec$ is consistently replaced by $\rvec'$), their explicit values naturally change. In contrast, the Laplacian of $\Lvec$ exhibits a more complex transformation, acquiring an additional term proportional to the Laplacian of the velocity field. This sensitivity of $\Lvec$ and its derivatives to the origin underscores the importance of clearly stating the chosen reference frame for consistent interpretation in physical analyses. For practical applications, selecting an origin judiciously (e.g., at a point of symmetry or a vortex core) can greatly simplify calculations and enhance the physical relevance of the results.

\subsection{Edicity: The Generalized Angular Momentum Energy}

In fluid dynamics, energy concepts provide fundamental insights into flow behavior. Analogous to kinetic energy and enstrophy, we can define an angular momentum energy density, $E_L$, based on the magnitude of $\Lvec$ as:

\begin{equation}
E_L = \frac{1}{2} \rho \Lvec \cdot \Lvec = \frac{1}{2} \rho |\Lvec|^2
\label{eq:EL-definition}
\end{equation}

We propose the name ``Edicity" for $E_L$ roughly based on ``Eddy Energy".\\

The units of $E_L$ are consistent with an energy density ($[M] [L]^{-1} [T]^{-2}$). To derive the evolution equation for $E_L$, we start from the transport equation for $\Lvec$ (Eq. \ref{eqn:angmom-full} from Section 2):

\[
\rho \frac{D \Lvec}{Dt} = -\rvec \times \nabla p + \mu (\nabla^2 \Lvec - 2 \omvec) 
\]

We take the dot product of this equation with $\Lvec$:

\[
\Lvec \cdot \left( \rho \frac{D \Lvec}{Dt} \right) = \Lvec \cdot \left( -\rvec \times \nabla p \right) + \Lvec \cdot \left( \mu (\nabla^2 \Lvec - 2 \omvec) \right)
\]

The left-hand side can be written in terms of $E_L$: $\rho \Lvec \cdot \frac{D \Lvec}{Dt} = \frac{D E_L}{Dt}$. Utilizing the vector identity $\mathbf{A} \cdot \nabla^2 \mathbf{A} = \frac{1}{2} \nabla^2 (\mathbf{A} \cdot \mathbf{A}) - |\nabla \mathbf{A}|^2$ for the viscous term, the evolution equation for $E_L$ becomes:

\begin{equation}
\frac{D E_L}{Dt} = -\Lvec \cdot (\rvec \times \nabla p) + \nu  \nabla^2 E_L - \mu \left[ |\nabla \Lvec|^2 + 2 \Lvec \cdot \omvec \right]
\label{eq:EL-evolution-final}
\end{equation}

This equation reveals the mechanisms governing the change of edicity which includes work by pressure torques and diffusion terms.  This equation has important applications in the study of turbulence which we will explore in a separate article.\\

A few remarks in the context of Eq.~\ref{eq:EL-evolution-final}.  In classical vortex dynamics, topological features are often characterized by the helicity density, defined as the projection of vorticity onto the velocity field ($H = \uvec \cdot \omvec$). Helicity provides a measure of the local alignment between the translational and rotational kinematics. Within the $\Lvec$-framework, a natural analogue arises through the scalar coupling of the angular momentum density and the vorticity field, $\Lvec \cdot \omvec$, which manifests as a direct source/sink term in the evolution Eq.~\ref{eq:EL-evolution-final} for Edicity ($E_L$).\\

By expanding this scalar coupling using the definition $\Lvec = \rvec \times  \uvec$, we reveal a fundamental connection to the Lamb vector ($\mathbf{l} = \omvec \times \uvec$). Applying the properties of the scalar triple product, the inner product of $\Lvec$ and $\omvec$ can be cyclically permuted:

\begin{equation}
\Lvec \cdot \omvec = (\rvec \times  \uvec) \cdot \omvec =  \rvec \cdot (\uvec \times \omvec)
\end{equation}

Recognizing that $\uvec \times \omvec = -\mathbf{l}$, this relationship simplifies to:

\begin{equation}
\Lvec \cdot \omvec = - \rvec \cdot \mathbf{l}
\end{equation}

This identity demonstrates that the scalar product $\Lvec \cdot \omvec$ is exactly proportional to the radial projection of the vortex force. \\

This formulation yields a particularly elegant geometric insight when applied to Beltrami flows. A pure Beltrami flow is defined as a state where the velocity and vorticity vectors are perfectly aligned everywhere, such that the Lamb vector vanishes ($\uvec \times \omvec = \mathbf{0}$). From the identity above, it immediately follows that for any Beltrami field:

\begin{equation}
    \Lvec \cdot \omvec = 0 \quad \implies \quad \Lvec \perp \omvec \quad \forall \; \rvec
\end{equation}

Consequently, while classical fluid dynamics views Beltrami flows as states of maximal alignment ($\uvec \parallel \omvec$), the $\Lvec$-framework dictates that they are states of strict geometric orthogonality ($\Lvec \perp \omvec$). Physically, this orthogonality condition dictates that the local spin-dissipative sink in the $E_L$ budget is entirely suppressed. For the study of isotropic turbulence, where Beltrami-like coherent structures correspond to regions where the nonlinear energy cascade is locally depleted, checking the orthogonality condition $\Lvec \cdot \omvec \to 0$ provides a novel, strictly geometric diagnostic for identifying these dynamic structures within the flow domain.

\section{Lift Generation in Inviscid Flows}

Having seen how lift generation works in fully viscous flows, we now turn our attention to inviscid flows, which provides the basis for essential theories of aerodynamics.  As is well known,  in inviscid flows, the boundary layers collapse into vorticity sheets.  We will show that these vorticity sheets act like singular sources of angular momentum $\Lvec$ that ``turn" the flow.  The lift can be calculated in terms of circulation which can be expressed in terms of $L_z$.\\

The Laplacian identity $\nabla^2 L_z = 2 \omega_z$ reveals the connection between angular momentem, the circulation $\Gamma = \oint_C \uvec \cdot d\mathbf{l}$ and  vorticity.\\

Using  $\omega_z = \Gamma \delta^{(2)}(\mathbf{r} - \mathbf{r}_0)$ at the effective vortex position $\mathbf{r}_0$ within the body, where $\delta^{(2)}$ is the 2D Dirac delta with units of inverse area we can write:

$$\nabla^2 L_z = 2 \Gamma \delta^{(2)}(\mathbf{r} - \mathbf{r}_0)$$

To evaluate the area integral $\int_A \nabla^2 L_z dA$, where $A$ is a suitably defined area, we also define $C = \partial A$ as the boundary contour and $d\mathbf{l} = \nvec \, dl$ with outward normal $\nvec$. The left-hand side is $\int_A \nabla^2 L_z \, dA = 2 \int_A \omega_z \, dA$. To account for the singularity at the origin, integrate over $A$ excluding a small disk $D_\epsilon$ of radius $\epsilon$ around the origin, then take the limit $\epsilon \to 0$:

$$\int_{A \setminus D_\epsilon} \nabla^2 L_z \, dA = \oint_C \nabla L_z \cdot d\mathbf{l} - \oint_{\partial D_\epsilon} \nabla L_z \cdot d\mathbf{l}$$

As $\epsilon \to 0$, the interior integral approaches 0 (since $\omega_z = 0$ outside the origin), so:
$$\oint_C \nabla L_z \cdot d\mathbf{l} = \lim_{\epsilon \to 0} \oint_{\partial D_\epsilon} \nabla L_z \cdot d\mathbf{l} = 2 \Gamma$$

confirming the singularity contributes the full source. This is analogous to Gauss's law in electrostatics, where the singularity (charge) sources the field flux, with the circulation $\Gamma$ acting as the ``charge" for the angular momentum ``potential" $L_z$.\\

Singularites like point vortices and vortex sheets represent angular momentum sources in an integral sense, acting as delta functions in the Laplacian relation $\nabla^2 L_z = 2 \omega_z$. \\

\begin{figure}[h!]
        \centering
        \subfloat[]{%
            \includegraphics[width=0.45\textwidth]{"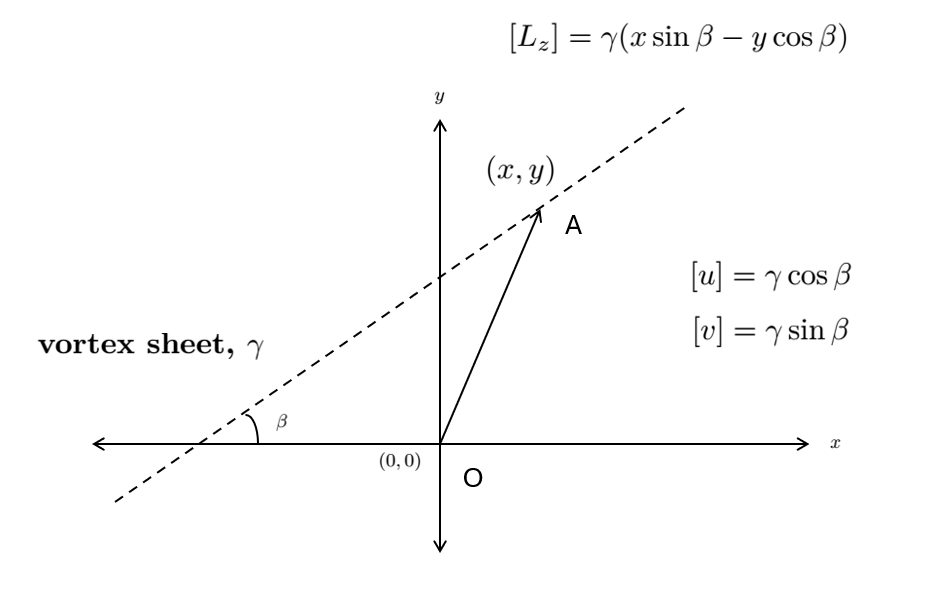"}\label{v1}}
        \subfloat[]{%
            \includegraphics[width=0.65\textwidth]{"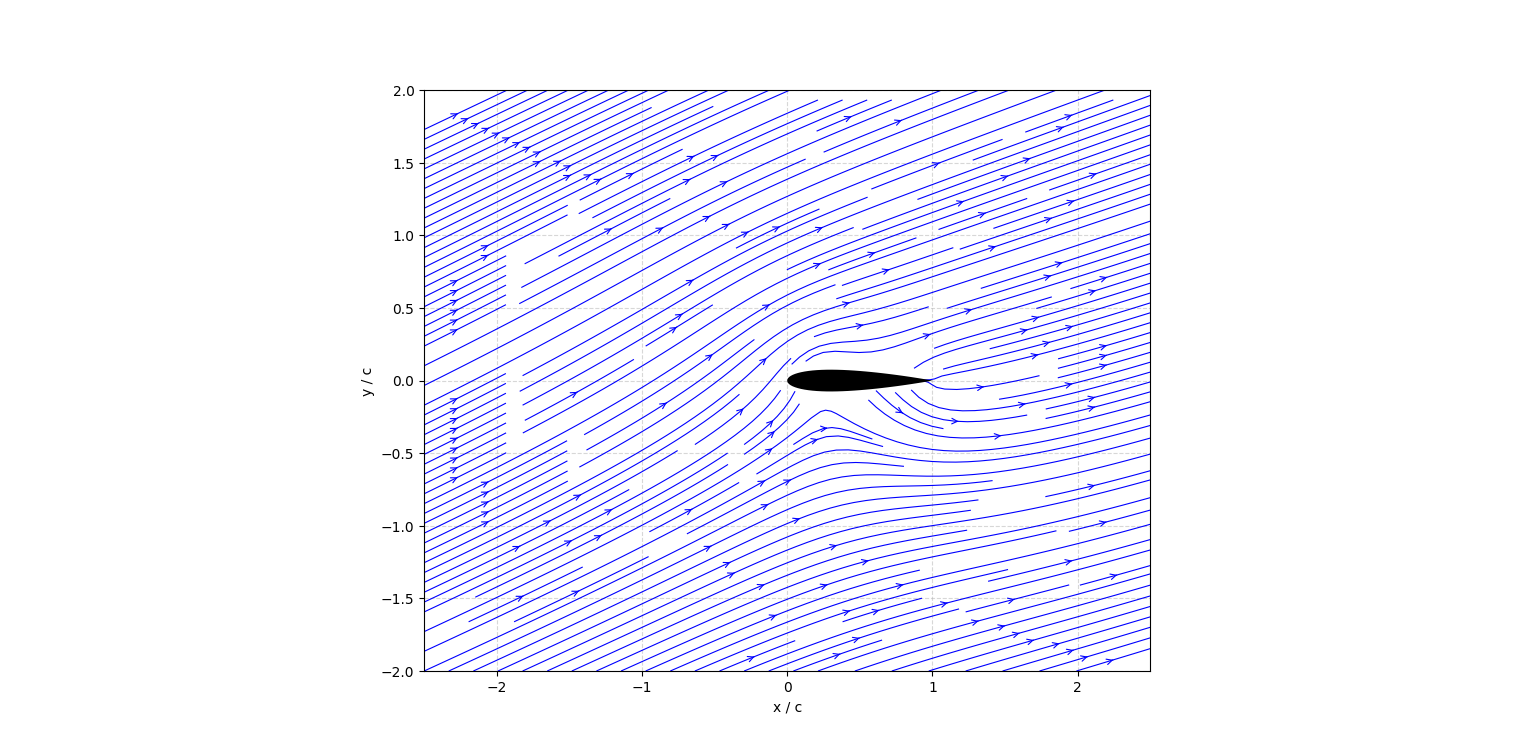"}\label{v2}}\\
            \caption{(a) The turning effect of a vortex sheet of strength $\gamma$ at angle $\beta$ to the frame of reference. (b) The turning effect of a vortex sheet of strength $\gamma(s)$ wrapped around an airfoil solved using panel methods.}       
        \label{vortices-1}
\end{figure}

The effect of other singular configurations of vorticity also results into singular distributions of $L_z$.  Consider a vortex sheet of constant strength $\gamma$ at an angle $\beta$ with the coordinate frame as shown in Figure \ref{vortices-1}(a). The velocity jump across the sheet is $[u_\gamma] = \gamma$.  Now since the sheet is at an angle, we can decompose the velocity jump into its $x$ and $y$ components as:

$$[u] = \gamma \cos \beta $$
$$[v] = \gamma \sin \beta.$$

The angular momentum $ L_z = x v - y u $, so the jump is:

\begin{equation}
[L_z] = x [v] - y [u] =  \gamma (x \sin \beta - y \cos \beta)
\label{eq:turning-sheet}
\end{equation}

Thus the vortex sheet acts as a singular source of angular momentum.  In this case since both $\beta$ and $\gamma$ are fixed, it imparts a constant angular momentum to flow along the sheet.  However consider Figure \ref{vortices-1}(b) which shows potential flow around a NACA 0015 airfoil at an angle of attack of $22^\circ$.  The flow has been modeled as a vortex sheet wrapped around the surface of the airfoil and solved by breaking it into a discrete number of ``panels".  The vortex sheet strength $\gamma(s)$ is not constant and varies from point to point along the length $s$ of the airfoil as does the angle $\beta$ which follows the shape of the airfoil. As can be seen the vortex sheet turns the flow as it crosses the airfoil, with the local turning effect governed by the local sheet strength $\gamma(s)$.  This offers a key insight into the lift generation mechanism: as the flow is turned downward, it generates a reaction on the airfoil which is exactly equal to the lift.  We will show the relationship between lift and angular momentum in the next section. \\

Before we move on, one final comment on compressible flows.  For compressible flows, it is well known that turning of flows is accomplished by oblique shocks.  In the section on compressible flow we will derive a relationship analogous to Eq. \ref{eq:turning-sheet} and show that oblique shocks also act as singular sources of angular momentum. \\

\subsection{The Kinematic Identity: Circulation and Angular Momentum}

We establish a direct, geometric identity between classical circulation and the angular momentum density field. This relationship serves as the translational bridge between traditional aerodynamic theory and the $\Lvec$-framework.\\

Consider an arbitrary two-dimensional flow field evaluated on a circular contour $C$ of radius $r$ centered at the origin. In polar coordinates $(r, \theta)$, the velocity vector is $\uvec = u_r \hat{e}_r + u_\theta \hat{e}_\theta$. The classical scalar circulation, $\Gamma$, is defined as the line integral of the velocity strictly tangent to the contour:

\begin{equation}
\Gamma = \oint_C \uvec \cdot d\mathbf{l} = \int_0^{2\pi} u_\theta \, r \, d\theta
\end{equation}

Within the same polar coordinate system, the out-of-plane angular momentum density, $\Lvec = \mathbf{r} \times \uvec$, possesses only a $z$-component. Because the cross product of the radial position vector $\mathbf{r} = r \hat{e}_r$ with the radial velocity $u_r \hat{e}_r$ vanishes, the angular momentum density is defined entirely by the tangential velocity:

\begin{equation}
L_z = r u_\theta
\end{equation}

Substituting this directly into the classical definition of circulation yields a kinematic identity:

\begin{equation}
\Gamma = \int_0^{2\pi} L_z \, d\theta
\end{equation}

This shows that any traditional aerodynamic theorem reliant on $\Gamma$ (such as the Kutta-Joukowski lift theorem) can be natively recast and evaluated using the local $\Lvec$ field.

\section{Hydrodynamic Impulse}

Hydrodynamic impulse formulations are useful for calculation of forces acting on bodies in motion (accelarting bodies or bodies with angular velocity) inside a fluid medium.  General references for Impulse formulations are Lamb \cite{lamb1924hydrodynamics}, Batchelor \cite{Batchelor2000}, Milne-Thompson \cite{MilneThomson1968}, Saffman \cite{Saffman1992}, Wu \cite{Wu2006}, etc.\\

As we will see in this section, formulation of the Impulse in the $\Lvec$ framework leads to surprisingly clean and powerful results with potential applications in aircraft and submarine design, fish locomotion etc.

\subsection{The Impulse Formulation in Terms of $\Lvec$}
The generalized hydrodynamic impulse $\Ivec$ for a localized vorticity distribution $\omvec$ within a fluid volume $V$ in $n$-dimensional space is defined as the first moment of the vorticity field, scaled by a dimensionality factor of $1/(d-1)$:

\begin{equation}
\Ivec = \frac{1}{d-1} \int_V \rvec \times \omvec \, dV
\end{equation}

To reformulate this volumetric integral in terms of the angular momentum density field, $\Lvec$, we recognize that the gradient of the scalar product $\rvec \cdot \uvec$ expands as\footnote{We utilize the vector identity for the gradient of a dot product: $\nabla (\mathbf{A} \cdot \mathbf{B}) = (\mathbf{A} \cdot \nabla)\mathbf{B} + (\mathbf{B} \cdot \nabla)\mathbf{A} + \mathbf{A} \times (\nabla \times \mathbf{B}) + \mathbf{B} \times (\nabla \times \mathbf{A})$.} $\nabla(\rvec \cdot \uvec) = (\rvec \cdot \nabla)\uvec + (\uvec \cdot \nabla)\rvec + \rvec \times (\nabla \times \uvec) + \uvec \times (\nabla \times \rvec)$. Noting that $\nabla \times \rvec = 0$, $(\uvec \cdot \nabla)\rvec = \uvec$, and $\nabla \times \uvec = \omvec$, this algebraically rearranges to:

\[ \rvec \times \omvec = \nabla(\rvec \cdot \uvec) - \uvec - (\rvec \cdot \nabla)\uvec \]

Next, we evaluate the curl of the angular momentum density, $\nabla \times \Lvec = \nabla \times (\rvec \times \uvec)$. Using the vector identity for the curl of a cross product, applying the incompressibility condition ($\nabla \cdot \uvec = 0$), and using the spatial dimension identity ($\nabla \cdot \rvec = d$), we find:
\[ \nabla \times \Lvec = -(\rvec \cdot \nabla)\uvec - (d-1)\uvec \]

Substituting this relation back into the expanded gradient equation allows us to eliminate the convective derivative $-(\rvec \cdot \nabla)\uvec$. This yields the generalized $n$-dimensional identity for the integrand:

\begin{equation}
\rvec \times \omvec = \nabla(\rvec \cdot \uvec) + (d-2)\uvec + \nabla \times \Lvec
\label{eq:impulse-identity}
\end{equation}

We can now integrate this expression over a control volume $V$ that encloses the solid body and extends to an outer boundary. Applying the divergence theorem to the first term on the right-hand side, and the vector curl theorem ($\int_V \nabla \times \Fvec \, dV = \oint_{\partial V} \nvec \times \Fvec \, dA$) to the third term, the volumetric impulse is transformed into a combination of a linear momentum integral and two surface fluxes:

\begin{equation}
\Ivec = \frac{d-2}{d-1}\int_V \uvec \, dV + \frac{1}{d-1}\oint_{\partial V} (\rvec \cdot \uvec)\nvec \, dA - \frac{1}{d-1}\oint_{\partial V} \Lvec \times \nvec \, dA
\label{eq:impulse-1}
\end{equation}

where $\partial V$ represents the bounding surfaces of the fluid domain, and $\nvec$ is the outward-pointing unit normal. Also, noting that $\nvec \times \Lvec = -\Lvec \times \nvec$. \\

In Eq. \ref{eq:impulse-1}, the classical impulse calculation is cleanly partitioned by dimensionality. It resolves into the pure linear inertia of the control volume (the volumetric impulse $\Ivec_{\text{vol}}$), the normal volumetric displacement of the fluid at the boundaries (the dilatational impulse $\Ivec_{\text{dil}}$), and the boundary flux of angular momentum ($\Lamvec$).

\subsection{Boundary Partitioning and the Solid Surface Fluxes}

In the derivation of the generalized impulse (Eq. \ref{eq:impulse-1}), extreme care must be exercised in the treatment of the boundary integral $\partial V$. For the practical aerodynamic problem of a body immersed in a fluid, the total boundary is partitioned into two distinct surfaces: $\partial V = S_\infty \cup S_b$, where $S_\infty$ is the far-field control surface and $S_b$ is the solid surface of the body. \\

If we define $\nvec$ as the standard outward-pointing normal vector from the solid body (meaning it points into the fluid), the normal for the fluid's inner boundary is $-\nvec$. The total boundary integral decomposes as:

\begin{equation}
\oint_{\partial V} (...) \, dA = \oint_{S_\infty} (...) \, dA - \oint_{S_b} (...) \, dA
\end{equation}

The evaluation of the inner boundary integrals on $S_b$ is strictly governed by the kinematic assumptions of the flow field—specifically, whether the fluid is viscous or inviscid. Assuming a body-fixed reference frame where the solid surface is stationary:\\

\textbf{1. The Viscous Limit (No-Slip Condition)} \\

For a real, viscous fluid, the boundary layer is attached to the wall, enforcing the no-slip and no-penetration conditions. The fluid velocity strictly vanishes at the solid boundary:

\begin{equation}
\uvec = \mathbf{0} \quad \text{on} \quad S_b
\end{equation}

Consequently, both the local dilatational scalar ($\rvec \cdot \uvec = 0$) and the local angular momentum density ($\Lvec = \mathbf{0}$) vanish entirely. For a viscous flow, the solid body contributes exactly zero to the surface flux terms in Eq. \ref{eq:impulse-1}, leaving the impulse budget to be dictated solely by the volume integral and the far-field fluxes at $S_\infty$.\\

\textbf{2. The Inviscid Limit (Slip Condition)} \\

For an ideal, inviscid fluid, the no-slip condition is relaxed. The flow cannot penetrate the solid geometry, but it is permitted to slip tangentially. The kinematic boundary condition is restricted to:

\begin{equation}
\uvec \cdot \nvec = 0 \quad \text{on} \quad S_b
\end{equation}

Because the tangential velocity ($\uvec_t$) survives on the boundary, the local angular momentum density is decidedly non-zero ($\Lvec = \rvec \times \uvec_t \neq \mathbf{0}$). Furthermore, unless the body geometry is a perfect sphere centered at the origin (where $\rvec$ is strictly parallel to $\nvec$), the position vector $\rvec$ will possess a tangential component, meaning the dilatational scalar ($\rvec \cdot \uvec$) also generally survives. \\

Therefore, in an inviscid flow, the solid surface $S_b$ actively carries non-zero boundary fluxes of angular momentum:

\begin{equation}
\Ivec_{boundary, S_b} = - \frac{1}{d-1}\oint_{S_b} (\rvec \cdot \uvec_t)\nvec \, dA + \frac{1}{d-1}\oint_{S_b} (\rvec \times \uvec_t) \times \nvec \, dA
\end{equation}

This distinction is mathematically vital: it demonstrates that even in a purely irrotational potential flow, the tangential slip velocity along the solid surface natively generates a bound angular momentum flux that must be resolved in the global force budget.

\subsection{The Angular Momentum Flux Vector, $\Lamvec$}

We define the final term in Eq. \ref{eq:impulse-1} as the angular momentum flux vector. Recalling the boundary partition $\partial V = S_\infty \cup S_b$, this flux must generally be evaluated over both the far-field outer control surface ($S_\infty$) and the solid body surface ($S_b$). We define the angular momentum flux vector across any arbitrary surface $S$ as:

\begin{equation}
\Lamvec(S) \equiv \oint_{S} (\Lvec \times \nvec) \, dA
\label{eq:Omega-def}
\end{equation}

Physically, $\Lamvec(S)$ quantifies the net outflow of angular momentum from the region bounded by $S$. For a stationary solid body immersed in a viscous fluid, the no-slip condition dictates that $\uvec = \mathbf{0}$ on the wall. Therefore, both the local angular momentum density ($\Lvec = \mathbf{0}$) and the dilatational scalar ($\rvec \cdot \uvec = 0$) vanish entirely on $S_b$. With the solid boundary contributions eliminated, the boundary integrals in Eq. \ref{eq:impulse-1} depend entirely on the outer control surface (which we will hereafter denote simply as $S$ for notational convenience). Consequently, evaluating $\Lamvec(S)$ provides a direct, integrated ``snapshot'' of exactly how much rotational debt the body has imparted to the surrounding fluid wake. Substituting Eq.~\ref{eq:Omega-def} into Eq.~\ref{eq:impulse-1} yields the generalized impulse budget:

\begin{equation}
\Ivec = \frac{d-2}{d-1}\int_V \uvec \, dV + \frac{1}{d-1}\oint_{S} (\rvec \cdot \uvec)\nvec \, dA - \frac{1}{d-1} \Lamvec(S)
\label{eq:impulse-2}
\end{equation}

To illustrate the geometric utility of this flux, consider a two-dimensional flow in the $xy$-plane, where $\uvec = u\ihat + v\jhat$ and $\rvec = x\ihat + y\jhat$. The angular momentum density is strictly out-of-plane, given by $\Lvec = L_z \khat$, where $L_z = xv - yu$. For a cylindrical control surface of unit depth bounded by a closed contour $C$, the differential area element is $dA = dl$. The outward unit normal vector to the contour is $\nvec = n_x\ihat + n_y\jhat$. Evaluating the cross product in the integrand yields:

\begin{equation}
\Lvec \times \nvec = (L_z \khat) \times (n_x\ihat + n_y\jhat) = L_z (-n_y\ihat + n_x\jhat)
\end{equation}

Traversing the contour $C$ in the standard counter-clockwise direction, the unit tangent vector is identically $\mathbf{t} = -n_y\ihat + n_x\jhat$. Substituting this geometric relationship back into the cross product, we find $\Lvec \times \nvec = L_z \mathbf{t}$. Recognizing that the unit tangent scaled by the differential arc length defines the vector path element ($d\mathbf{l} = \mathbf{t} \, dl$), the 2D angular momentum flux elegantly reduces to:

\begin{equation}
\Lamvec = \oint_C L_z \, d\mathbf{l}
\end{equation}

This formulation provides a striking structural analogue to the classical definition of circulation:

$$\Gamma = \oint_C \uvec \cdot d\mathbf{l}$$

However, while $\Gamma$ is a scalar metric representing the total macroscopic spin, $\Lamvec(S)$ is a vector quantity. Because $\Lamvec$ integrates the scalar field $L_z$ (which natively encodes the spatial distribution of the momentum) against the vector path $d\mathbf{l}$, it seamlessly resolves the flow's asymmetry into distinct Cartesian force components. This preservation of directional information makes it a significantly more powerful diagnostic tool for non-symmetric or complex maneuvering bodies.\\

Furthermore, traditional potential formulations (such as the Kutta-Joukowski theorem) rely heavily on irrotational assumptions and become invalid in the presence of viscosity and flow separation. In contrast, $\Lamvec(S)$ provides a deterministic, kinematic pathway to compute forces directly from the velocity fields of fully viscous, separated flows. It bypasses the requirement for irrotationality, allowing the circulatory and viscous forces to be evaluated with the exact same mathematical rigor utilized for ideal fluids.

\subsection{Derivation of the Force}

We derive an exact expression for the aerodynamic force $\Fvec$ acting on a body in an incompressible flow.  The derivation is based on the material derivative of the hydrodynamic impulse $\Ivec$, which is a classical result (see, e.g., Saffman \cite{Saffman1992}, Wu \cite{wu1981theory}).  Let $V$ be a fixed control volume that encapsulates the fluid domain, with boundary $S = \partial V$ composed of the solid body surface $S_b$ and an outer surface $S_\infty$ far from the body.  For a general $n$-dimensional space, the force on the body is given by

\begin{equation}
\Fvec = \rho \; \frac{D\Ivec}{Dt}; \qquad 
\Ivec = \frac{1}{d-1} \int_V (\rvec \times \omvec) \, dV
\label{eq:force-impulse-exact}
\end{equation}

where $\omvec = \nabla \times \uvec$ is the vorticity, and $D/Dt$ denotes the material derivative. Equation~\eqref{eq:force-impulse-exact} holds for any incompressible flow in an unbounded domain with constant pressure at infinity, regardless of viscosity or the presence of wakes.\footnote{Eq.~\ref{eq:force-impulse-exact} is exact for infinite domains. But for calculation in finite domians, some surface integrals which may not decay fast enough may need to be included in Eq.~\ref{eq:force-impulse-exact}.  In practice, for high Reynolds number flow, especially for CFD related calculations, the domain bounding box is made large enough to ``simulate" an infinite domain and keep the surface effects within the limits of the overall errors in the calculation.  Hence one maybe be justified in using the infinite domain approximation here.}  Using the generalized $n$-dimensional vector identity Eq.~\ref{eq:impulse-identity} restated below:

\begin{equation}
\rvec \times \omvec = \nabla(\rvec \cdot \uvec) + (d-2)\uvec + \nabla \times \Lvec
\end{equation}

Integrating over $V$ and converting to surface integrals by Gauss's theorem and the vector curl theorem ($\int_V \nabla \times \Lvec \, dV = \oint_S \nvec \times \Lvec \, dA$), we obtain the generalized impulse decomposition:

\begin{equation}
\Ivec = \frac{1}{d-1}\underbrace{\oint_{S} (\rvec \cdot \uvec) \nvec \, dA}_{\Ivec_{\text{dil}}} \;+\; \frac{d-2}{d-1}\underbrace{\int_V \uvec \, dV}_{\Ivec_{\text{vol}}} \;+\; \frac{1}{d-1}\underbrace{\oint_{S} (\nvec \times \Lvec) \, dA}_{\boldsymbol{\Omega}} 
\label{eq:impulse-decomposition-exact}
\end{equation}

This decomposition is exact for any incompressible flow. The three foundational geometric integrals are defined as: $\Ivec_{\text{dil}}$ which we term the dilatational impulse (normal displacement of fluid on the boundary), $\Ivec_{\text{vol}}$ which we term the volumetric impulse (total linear momentum inside $V$), and $\boldsymbol{\Omega}$ which we term the rotational flux (representing the boundary flux of angular momentum, noting that $\nvec \times \Lvec = -\Lamvec$). \\

Recalling our boundary partition $S = S_\infty \cup S_b$, the evaluation of the surface integrals ($\Ivec_{\text{dil}}$ and $\boldsymbol{\Omega}$) on the solid body depends strictly on the kinematic assumptions of the flow. In a viscous flow, the no-slip condition ($\uvec = \mathbf{0}$) dictates that both surface terms vanish entirely on $S_b$. In an inviscid flow, tangential slip velocities survive, meaning $S_b$ actively carries bound fluxes of dilatation and angular momentum. \\

Since $V$ is fixed, the material derivative of $\Ivec$ can be expanded using the Reynolds transport theorem by breaking down the total derivative into a local temporal variation $\frac{\partial}{\partial t}$ and a convective flux on the boundary (scaling naturally by the $n$-dimensional prefactor): 

\begin{equation}
\frac{D\Ivec}{Dt} = \frac{\partial \Ivec}{\partial t} + \frac{1}{d-1}\oint_{S} (\rvec \times \omvec) (\uvec \cdot \nvec) \, dA 
\label{eq:rtt-exact}
\end{equation}

Because the solid body is impermeable, the relative normal velocity on the boundary vanishes ($\uvec \cdot \nvec = 0$ on $S_b$). Consequently, the convective flux of vorticity across the solid body is mathematically annihilated for both viscous and inviscid flows. This flux term evaluates solely over the outer boundary $S_\infty$. Substituting the decomposition \eqref{eq:impulse-decomposition-exact} into \eqref{eq:rtt-exact} and multiplying by the density $\rho$ gives the exact force formula:

\begin{equation}
\Fvec = \frac{\rho}{d-1} \frac{\partial}{\partial t} \bigl( \Ivec_{\text{dil}} + (d-2)\Ivec_{\text{vol}} + \boldsymbol{\Omega} \bigr) \;+\; \frac{\rho}{d-1} \oint_{S_\infty} (\rvec \times \omvec) (\uvec \cdot \nvec) \, dA 
\label{eq:force-exact-raw}
\end{equation}

Eq.~\ref{eq:force-exact-raw} is the most general force expression.  It is valid for any incompressible flow (viscous or inviscid, steady or unsteady) and any choice of control volume $V$ that contains the body and all vorticity.\\

We can now separate the total force into three physically distinct contributions that arise naturally from the structure of Eq.~\ref{eq:force-exact-raw}:

\begin{equation}
\Fvec = \Fvec_{\text{nc}} + \Fvec_{\text{ca}} + \Fvec_{\text{comb}} 
\label{eq:force-exact-components}
\end{equation}

where the non‑circulatory added mass force $\Fvec_{\text{nc}}$ originates from the time variation of the displacement‑related impulses.  It represents the inertial force required to accelerate the fluid that is simply displaced by the body, even in the absence of vorticity.  By definition:

\begin{equation}
\Fvec_{\text{nc}} \equiv \frac{\rho}{d-1} \frac{\partial}{\partial t} \bigl( \Ivec_{\text{dil}} + (d-2)\Ivec_{\text{vol}} \bigr)
\end{equation}

The circulatory added mass force $\Fvec_{\text{ca}}$ arises from the time variation of the rotational flux $\boldsymbol{\Omega}$.  It captures the force due to the unsteady growth or decay of the body's rotational debt (e.g., the change of bound circulation or the evolution of the wake).  It is defined as:

\begin{equation}
\Fvec_{\text{ca}} \equiv \frac{\rho}{d-1} \frac{\partial \boldsymbol{\Omega}}{\partial t}
\end{equation}

The Combination term (lift/drag from advection of vorticity moment) $\Fvec_{\text{comb}}$ is the convective flux of the vorticity moment across the far-field control surface. It accounts for the steady (or unsteady) advection of rotational structures out of the control volume, and in classical steady‑state aerodynamics it reduces to the Kutta–Joukowski lift.  Because the solid surface flux is identically zero, it is defined solely on $S_\infty$:

\begin{equation}
\Fvec_{\text{comb}} \equiv \frac{\rho}{d-1} \oint_{S_\infty} (\rvec \times \omvec) (\uvec \cdot \nvec) \, dA 
\end{equation}

Equations~\eqref{eq:force-exact-raw} and \eqref{eq:force-exact-components} are exact and involve no approximations\footnote{See \S 6.5 below for comparison to the formalisms of Wu and Noca.}.  They hold for any incompressible flow, whether viscous or inviscid, attached or separated, steady or unsteady.  The only requirement is that the control volume $V$ encloses the body and all vorticity, and that the integrals converge (which is guaranteed for flows with finite vorticity support or sufficiently rapid decay).\\

Each term in the decomposition depends on the choice of origin through $\rvec$ in $\Lvec$ and in the definition of $\Ivec_{\text{dil}}$.  However, the total force $\Fvec$ is origin‑independent.  The origin dependence cancels among $\Fvec_{\text{nc}}$, $\Fvec_{\text{ca}}$, and $\Fvec_{\text{comb}}$ when they are summed.  This cancellation is a powerful consistency check of the framework.\\

A key advantage of the $\Lvec$‑based triple decomposition is that the non‑circulatory force is defined solely from the instantaneous velocity field $\uvec$, without any assumption of irrotationality or potential flow.  Consequently, $\Fvec_{\text{nc}}$ automatically incorporates the influence of viscosity, boundary layers, and separated wakes on the fluid's inertial resistance to the body's acceleration.  We therefore refer to $\Fvec_{\text{nc}}$ as the exact viscous added mass force.\\

In the classical potential‑flow limit, $\Fvec_{\text{nc}}$ recovers the standard added mass coefficients (e.g., $\frac{2}{3}\pi\rho R^3\dot{U}$ for a sphere as shown in an example below).  However, when the flow is viscous or contains vorticity, $\Fvec_{\text{nc}}$ remains well‑defined and accounts for the effective ``entrained'' fluid mass that must be accelerated, including the displacement thickness of boundary layers and trapped wake regions.  This makes the $\Lvec$ framework particularly suitable for unsteady maneuvers in real fluids, where added mass is not a constant geometric property but evolves with the flow state.

\subsection{Discussion}

In generalized unsteady aerodynamics, the total force on a body is classically partitioned into an inertial fluid reaction (the added mass) and a convective rotational force (the circulatory lift). To evaluate these forces rigorously from the vorticity field, the exact formalisms of Wu \cite{wu1981theory} and Noca \cite{noca1999comparison} utilize vector identities to map the fluid acceleration entirely to the total time derivative of the hydrodynamic impulse. \\

Noca’s generalized impulse equation expresses the exact aerodynamic force on a body within a finite control volume $V$ as:

\begin{equation}
\mathbf{F} = -\frac{\rho}{d-1} \frac{d}{dt} \int_V (\mathbf{x} \times \boldsymbol{\omega}) \, dV + \rho \oint_S \left[ \frac{1}{2} u^2 \mathbf{n} + (\mathbf{x} \cdot \dot{\mathbf{u}})\mathbf{n} - \mathbf{x}(\mathbf{n} \cdot \dot{\mathbf{u}}) + \mathbf{u}(\mathbf{u} \cdot \mathbf{n}) - (\mathbf{n} \cdot \boldsymbol{\omega} \times \mathbf{u})\mathbf{x} \right] dA \label{eq:noca_exact}
\end{equation}

where $d$ is the spatial dimension. In the limit of an infinite domain ($S \to \infty$), the irrotational far-field boundary fluxes vanish. The equation recovers Wu’s foundational result, where the unsteady force is dictated solely by the total derivative of the fluid impulse, $\mathbf{I}$:

\begin{equation}
\mathbf{F}_{\text{Wu}} = -\frac{\rho}{d-1} \frac{d}{dt} \int_{V_\infty} (\mathbf{x} \times \boldsymbol{\omega}) \, dV \equiv \frac{d\mathbf{I}}{dt} 
\label{eq:wu_exact}
\end{equation}

While mathematically elegant, these classical formulations fundamentally do not explicitly identify  the added mass. The volume integral in Eq.~\ref{eq:wu_exact} encompasses all vorticity within the fluid domain, which must be linearly decomposed into the bound circulation attached to the body ($\boldsymbol{\omega}_{\text{bound}}$) and the free vorticity shedding into the wake ($\boldsymbol{\omega}_{\text{wake}}$):

\begin{equation}
\mathbf{F}_{\text{Wu}} = \frac{d}{dt} \mathbf{I}_{\text{bound}} + \frac{d}{dt} \mathbf{I}_{\text{wake}}
\end{equation}

By applying a single, total time derivative ($d/dt$) to this global integral, the formalism mathematically fuses two entirely distinct physical mechanisms: (i) conservative inertia (added mass): the reversible, inertial force required to accelerate the irrotational fluid mass surrounding the moving body. (ii) non-conservative convection (circulatory lift): the dissipative force fluctuation induced by the shedding and subsequent downstream advection of new vorticity (e.g., the Wagner/Theodorsen effects).\\

Contrast this with Eq.~\ref{eq:force-exact-raw} where it is possible to cleanly segregate the added mass terms.

\subsection{Example 1: The Accelerating Cylinder}

To validate the force equation (Eq. \ref{eq:force-exact-raw}) in two dimensions ($d=2$), we compute the exact non-circulatory force budget for a circular cylinder of radius $R$ accelerating with instantaneous velocity $U(t) \ihat$. We can bypass the solid boundary integrals ($S_b$) by employing classical impulse theory. We define the control volume $V$ to encompass the entire spatial domain out to the far-field boundary $S_\infty$, treating the interior of the solid body as fictitious fluid moving uniformly at $U(t) \ihat$. This embeds the physical solid boundary as an internal vortex sheet, leaving $S_\infty$ as the sole topological boundary of the system.\\

The potential flow field in the fluid domain ($r > R$) is governed by the 2D dipole potential $\phi = -U(t) \frac{R^2}{r} \cos \theta$, yielding the velocity components:

\begin{equation}
u_r = U(t) \frac{R^2}{r^2} \cos \theta, \quad u_\theta = U(t) \frac{R^2}{r^2} \sin \theta
\end{equation}

For applying Eq.~\ref{eq:force-exact-raw} we first note that the last integral is zero for potential flows since $\omvec =0$.  We are  therefore required only to evaluate the three momentum budget terms. We begin with the volumetric impulse, $\Ivec_\text{vol}$. For a 2D system ($d=2$), the geometric prefactor $(d-2)$ is identically zero and hence the term drops off. \\

Evaluating the normal boundary flux at the far-field boundary $S_\infty$ (where $r = r_\infty$ and $\nvec = \evec_r$), the dilatational impulse resolves to:

\begin{equation}
\Ivec_{\text{dil}, x}(S_\infty) = \oint_{S_\infty} (\rvec \cdot \uvec) n_x \, dA = \int_0^{2\pi} \left( r_\infty U \frac{R^2}{r_\infty^2} \cos \theta \right) \cos \theta \, (r_\infty d\theta) = \pi R^2 U(t)
\end{equation}

To evaluate the rotational flux $\boldsymbol{\Omega}$  at infinity we note that while the flow is purely irrotational ($\omvec = \mathbf{0}$), the angular momentum density $\Lvec$ at the boundary is non-zero. Computing $\Lvec = \rvec \times \uvec = (r \evec_r) \times (u_r \evec_r + u_\theta \evec_\theta) = r u_\theta \khat$. Taking the cross product with the outward normal vector yields $\nvec \times \Lvec = \evec_r \times (r u_\theta \khat) = -r u_\theta \evec_\theta$. The $x$-component is defined by the projection $\evec_\theta \cdot \ihat = -\sin\theta$. Integrating this over $S_\infty$:

\begin{equation}
\Omega_x(S_\infty) = \oint_{S_\infty} (\nvec \times \Lvec)_x \, dA = \int_0^{2\pi} \left( -r_\infty U \frac{R^2}{r_\infty^2} \sin \theta \right) (-\sin \theta) \, (r_\infty d\theta) = \pi R^2 U(t)
\end{equation}

It is easy to see (by a symmetry argument or evaluation of the integrals) that $\Omega_y(S_\infty) = \Ivec_{\text{dil}, y} = 0 $. Applying the generalized $1/(d-1)$ Eq.~\ref{eq:force-exact-raw}  for $d=2$, the total force required to accelerate the global momentum of the system relies entirely on the two mathematically equal boundary fluxes:

\begin{equation}
F_{\text{total}} =  \frac{d}{dt} \bigl( \Ivec_{\text{dil}, x}(S_\infty) + \Omega_x(S_\infty) \bigr) = 2 \rho \pi R^2 \dot{U}
\end{equation}

Because this global formulation accounts for the acceleration of the entire domain including the fictitious fluid inside the body, the true, purely aerodynamic non-circulatory force ($F_{nc}$) exerted on the solid cylinder is found by subtracting the inertial force required to accelerate that internal fictitious fluid ($\rho \pi R^2 \dot{U}$):

\begin{equation}
F_{nc} = F_{\text{total}} - \rho \pi R^2 \dot{U} = 2 \rho \pi R^2 \dot{U} - \rho \pi R^2 \dot{U} = \rho \pi R^2 \dot{U}
\end{equation}

This demonstrates the application of Eq. \ref{eq:force-exact-raw} to recover the classical added mass for a 2D cylinder ($m_a = \rho \pi R^2$). It proves that the $\Lvec$-framework provides a continuous, unified global momentum budget across dimensionalities, capturing the inertial force through the strict geometric interplay of the rotational and dilatational far-field fluxes.

\subsection{Example 2: The Accelerating 3D Sphere}

To validate the force equation (Eq. \ref{eq:force-exact-raw}) in three dimensions ($d=3$), we compute the exact non-circulatory force budget for a sphere of radius $R$ accelerating with instantaneous velocity $U(t) \khat$ in an unbounded fluid.  We follow the same procedure as in the 2D case, but with the notable difference that here $\Ivec_\text{vol}$ also contributes.\\

To capture the total hydrodynamic impulse of the system while avoiding the evaluation of slip-fluxes on the solid body, we again employ the same fictitious fluid methodology used in Example 1. We define the control volume $V$ to encompass the entire spatial domain out to the far-field boundary $S_\infty$, treating the interior of the solid body as fictitious fluid moving uniformly at $U(t) \khat$. This topologically embeds $S_b$ as an internal interface, leaving the budget strictly dependent on $V$ and $S_\infty$.\\

The potential flow field in the fluid domain ($r > R$) is governed by the 3D dipole potential $\phi = -U(t) \frac{R^3}{2r^2} \cos \theta$, yielding the velocity components:

\begin{equation}
u_r = U(t) \frac{R^3}{r^3} \cos \theta, \quad u_\theta = U(t) \frac{R^3}{2r^3} \sin \theta
\end{equation}

Applying Eq. \ref{eq:force-exact-raw}  requires evaluating the three momentum budget terms. We begin with the volumetric impulse, $\Ivec_\text{vol}$. A fundamental property of the 3D dipole field is that its integrated linear momentum over the unbounded fluid domain is exactly zero. Thus, the total linear momentum is entirely contained within the fictitious fluid of the body interior:

\begin{equation}
\Ivec_{\text{vol}, z} = \int_{fluid} u_z \, dV + \int_{body} U \, dV = \frac{4}{3} \pi R^3 U(t)
\end{equation}

where the first integral is zero.  For $d=3$, the volumetric coefficient is exactly $(3-2)=1$. The fictitious body momentum is fully retained in the budget.\\

Evaluating the normal boundary flux at the far-field boundary $S_\infty$ (where $r = r_\infty$, $\nvec = \evec_r$, and $dA = r_\infty^2 \sin \theta \, d\theta d\varphi$, $n_z = \evec_r \cdot \khat = \cos \theta$), the dilatational impulse resolves to:

\begin{equation}
\Ivec_{\text{dil}, z}(S_\infty) = \oint_{S_\infty} (\rvec \cdot \uvec) n_z \, dA = \int_0^{2\pi} \int_0^\pi \left( r_\infty U \frac{R^3}{r_\infty^3} \cos \theta \right) \cos \theta \left( r_\infty^2 \sin \theta \right) \, d\theta d\varphi = \frac{4}{3} \pi R^3 U(t)
\end{equation}

We must also evaluate the rotational flux $\boldsymbol{\Omega}$ at infinity. The angular momentum density is $\Lvec = \rvec \times \uvec = (r \evec_r) \times (u_r \evec_r + u_\theta \evec_\theta) = r u_\theta \evec_\varphi$. Taking the cross product with the outward normal vector yields $\nvec \times \Lvec = \evec_r \times (r u_\theta \evec_\varphi) = -r u_\theta \evec_\theta$. The $z$-component is defined by the projection $\evec_\theta \cdot \khat = -\sin\theta$. Integrating this over $S_\infty$:

\begin{equation}
\Omega_z(S_\infty) = \oint_{S_\infty} (\nvec \times \Lvec)_z \, dA = \int_0^{2\pi} \int_0^\pi \left( -r_\infty U \frac{R^3}{2 r_\infty^3} \sin \theta \right) (-\sin \theta) \left( r_\infty^2 \sin \theta \right) \, d\theta d\varphi = \frac{4}{3} \pi R^3 U(t)
\end{equation}

Applying the force equation (Eq. \ref{eq:force-exact-raw}) with the $1/(d-1)$ prefactor for $d=3$, the total force required to accelerate the global momentum of the system relies on an exact equipartition of the three impulses:

\begin{equation}
F_{\text{total}} = \frac{\rho}{2} \frac{d}{dt} \left( \frac{4}{3} + \frac{4}{3} + \frac{4}{3} \right) \pi R^3 U = 2 \rho \pi R^3 \dot{U}
\end{equation}

Because this global formulation accounts for the acceleration of the fictitious fluid inside the body, the true purely aerodynamic non-circulatory force ($F_{nc}$) exerted on the solid sphere is found by subtracting the displacement mass force ($\frac{4}{3} \rho \pi R^3 \dot{U}$) required to accelerate that internal fictitious fluid:

\begin{equation}
F_{nc} = F_{\text{total}} - \frac{4}{3} \rho \pi R^3 \dot{U} = \frac{2}{3} \rho \pi R^3 \dot{U}
\end{equation}

This rigorous application of Eq.~\ref{eq:force-exact-raw} exactly recovers the classical added mass for a sphere ($m_a = \frac{2}{3} \rho \pi R^3$), proving that the $\Lvec$-framework is perfectly consistent across dimensionalities when the global boundary fluxes are mathematically maintained.

\subsection{Example 3: Numerical Evaluation of Added Mass from 3D RANS Data}

The real value of the generalized impulse framework via Eq.~\ref{eq:force-exact-raw} is the ability to compute the added mass for fully viscous flows.\\

We now develop a highly robust methodology for extracting the instantaneous non-circulatory (added mass) force directly from discrete computational fluid dynamics (CFD) data. We assume a 3D Reynolds-Averaged Navier-Stokes (RANS) simulation provides a time-resolved velocity field $\bar{\uvec}(\rvec, t)$ within a computational domain containing a submarine of volume $V_{sub}$ moving with kinematic velocity $\mathbf{U}(t)$. \\

Because the RANS solver enforces the viscous no-slip condition on the hull, direct surface integration of the hydrodynamic impulse on the complex body geometry is mathematically invalid (as the boundary fluxes identically vanish). To evaluate the exact inertial force without relying on noisy pressure field integrations, we employ the discrete fictitious fluid approach.\\

We define an arbitrary, stationary rectangular bounding box within the fluid domain that completely encapsulates the submarine. The volume of this box is $V_{box}$ and its external boundary is $S_{box}$. \\

To mathematically erase the solid submarine boundary from the topological budget, we apply a fictitious fluid mask. We identify all computational grid cells that fall physically inside the submarine hull. The velocity of these internal cells is overwritten with the instantaneous rigid-body velocity of the submarine, while the exterior cells retain the RANS velocity field:

\begin{equation}
\uvec^*(\rvec, t) = 
\begin{cases} 
\mathbf{U}(t) & \text{if } \rvec \in V_{sub} \\
\bar{\uvec}(\rvec, t) & \text{if } \rvec \notin V_{sub} 
\end{cases}
\end{equation}

With the internal boundary condition eliminated, the non-circulatory force depends strictly on the volumetric inertia of the masked domain and the dilatational flux across the simple, flat boundary $S_{box}$. \\

For a discrete grid with cell volumes $\Delta V_i$ and boundary face areas $\Delta A_j$ (with outward normal $\nvec_j$), we compute the instantaneous volumetric and dilatational impulses at each time step:

\begin{align}
\Ivec_{\text{vol}}(t) &= \sum_{i \in V_{box}} \uvec^*_i(t) \, \Delta V_i \\
\Ivec_{\text{dil}}(t) &= \sum_{j \in S_{box}} \left( \rvec_j \cdot \uvec^*_j(t) \right) \nvec_j \, \Delta A_j
\end{align}

Because the added mass is an inertial phenomenon dominated by the large-scale fluid displacement, the use of the Reynolds-averaged field ($\bar{\uvec}$) perfectly captures the macroscopic displacement impulses without requiring sub-grid scale modeling.\\

Using the $d=3$ in the fully viscous equation (derived from Eq. \ref{eq:force-exact-raw}), the total global non-circulatory force required to accelerate the masked system is the temporal derivative of the combined impulses. Applying a discrete finite difference operator (e.g., a central difference scheme over discrete time steps $t_k$):

\begin{equation}
\Fvec_{\text{nc, total}}(t_k) = \frac{\rho}{2} \frac{d}{dt} \Big[ \Ivec_{\text{dil}}(t_k) + \Ivec_{\text{vol}}(t_k) \Big] \approx \frac{\rho}{4 \Delta t} \Big[ \bigl( \Ivec_{\text{dil}} + \Ivec_{\text{vol}} \bigr)_{k+1} - \bigl( \Ivec_{\text{dil}} + \Ivec_{\text{vol}} \bigr)_{k-1} \Big]
\end{equation}

The force calculated above represents the effort required to accelerate both the external fluid and the internal fictitious fluid. To isolate the true aerodynamic added mass force exerted strictly on the solid submarine hull, we subtract the trivial rigid-body inertia of the fictitious internal mass:

\begin{equation}
\Fvec_{\text{nc, aero}}(t_k) = \Fvec_{\text{nc, total}}(t_k) - \rho V_{sub} \dot{\mathbf{U}}(t_k)
\end{equation}

This numerical procedure allows the exact, instantaneous added mass tensor of a complex maneuvering vehicle to be extracted natively from the kinematics of a viscous solver, entirely bypassing the integration of surface stress tensors over the hull.\\

It may be noted that other terms in Eq.~\ref{eq:force-exact-raw} may also be similarly evaluated by using quadratures for the surface integrals to yield $\Fvec_\text{ca}$ and $\Fvec_\text{comb}$.

\section{Geophysical Flows}

Geophysical flows, such as those in the Earth's atmosphere and oceans, are directly influenced by the planet's rotation, which imparts a natural angular momentum to fluid parcels. The angular momentum framework, with $\Lvec = \rvec \times \uvec$ defined relative to the planetary axis, is particularly well-suited for analyzing these systems.\\

Unlike vorticity, which is a local quantity requiring an artificial decomposition into planetary ($f$) and relative ($\zeta$) components to account for rotation, $\Lvec$ inherently incorporates the global rotational effects through torques. This non-local perspective avoids such constructs, providing a unified view where planetary rotation emerges naturally as a torque term, aligning with the conservation of absolute angular momentum in rotating frames.\\

Pedlosky \cite{pedlosky2013geophysical}, Vallis \cite{vallis2017atmospheric}, Olbers \cite{olbers2012ocean}, Holton \cite{holton2013introduction} are well known references for Geophysical Flows.

\subsection{Formation of Jet Streams via Conservation of Angular Momentum}

Jet streams are narrow bands of strong westerly winds in the upper atmosphere. Their formation involves conservation of angular momentum in meridional circulations like the Hadley cell. The $\Lvec$ framework naturally captures this, as $\Lvec$ directly quantifies the conserved quantity without approximations.\\

Let's start with the zonal (latitudinal) component of the Navier-Stokes equation in the rotating frame in the $(\lambda, \phi,z)$ ($\lambda$ - zonal, $\phi$-azimuthal, $z$-vertical up, with $a$ being the Earths radius and $\Omega_p$ the Earth's rotational angular velocity) coordinates:  

\begin{equation}
\frac{\partial u}{\partial t} + \uvec \cdot \nabla u - f v - \frac{u v \tan \phi}{a} = -\frac{1}{\rho} \frac{1}{a \cos \phi} \frac{\partial p}{\partial \lambda} + F_u
\label{eq:zonal-1}
\end{equation}

where $ f = 2 \Omega_p \sin \phi $ is Coriolis\footnote{With apologies to the reader, we use $\Omega_p$ for planetary spin although we have used $\Lamvec$ for circulatory impulse before.  But it should not cause any confusion.} parameter, $ -f v $ Coriolis term due to meridional velocity $v$, $\frac{u v \tan \phi}{a} $ is a metric term from spherical geometry (curvature effect), the Coriolis term $ -f v = -2 \Omega_p v \sin \phi  $ explicitly appears here, representing the deflection of the zonal flow due to Earth's rotation acting on meridional motion. $F_u$ is the friction term in the equation. \\

We will define an angular momentum  $m$ which we note is made of two components:  the first, arises from $u$, the relative zonal velocity (in the rotating frame), and the second is $\Omega_p a \cos \phi $, the planetary contribution from Earth's rotation at latitude $ \phi $, with $ a \cos \phi $ being the distance from the rotation axis. Using $a \cos \phi$ as the lever arm for angular momentum about the axis gives: 

\begin{equation}
m = (u + \Omega_p a \cos \phi) a \cos \phi
\label{eq:m-def}
\end{equation}

To derive the transport equation for $m$, we start with the zonal momentum Eq. \ref{eq:zonal-1} and transform it to express the evolution of $ m $. Rewrite the momentum equation in full form  as:

\begin{equation}
\frac{\partial u}{\partial t} + u \frac{1}{a \cos \phi} \frac{\partial u}{\partial \lambda} + v \frac{1}{a} \frac{\partial u}{\partial \phi} + w \frac{\partial u}{\partial z} - 2 \Omega_p \sin \phi v + \frac{u v \tan \phi}{a} = -\frac{1}{\rho} \frac{1}{a \cos \phi} \frac{\partial p}{\partial \lambda} + F_u
\label{eq:zonal-2}
\end{equation}

Rearrange Eq. \ref{eq:m-def} as:

\begin{equation}
u = \frac{m - \Omega_p a^2 \cos^2 \phi}{a \cos \phi}
\end{equation}

The material derivative of $m$ is given by:

\begin{equation}
\frac{D m}{Dt} = \frac{D}{Dt} \left( u a \cos \phi + \Omega_p a^2 \cos^2 \phi \right)
\end{equation}

For the first term above, since $ a $ is constant and also using \footnote{To see this, consider that $\frac{D (a\phi)}{Dt} = v$.} $\frac{D \phi}{Dt} = \frac{v}{a}$, we can show that:

$$a \cos \phi \frac{D u}{Dt} - u \sin \phi v$$

For the second term (the planetary part) we get:

$$\frac{D}{Dt} (\Omega_p a^2 \cos^2 \phi) =  -2 \Omega_p a v \cos \phi \sin \phi $$

So we have:

\begin{equation}
\frac{D m}{Dt} = a \cos \phi \frac{D u}{Dt} - u \sin \phi v - 2 \Omega_p a \cos \phi \sin \phi v
\label{eq:interm-1}
\end{equation}

Take $\frac{D u}{Dt}$ from the momentum Eq. \ref{eq:zonal-2} and substitute into Eq. \ref{eq:interm-1} and simplify to give:

\begin{equation}
\frac{D m}{Dt} = -\frac{1}{\rho} \frac{\partial p}{\partial \lambda} + a \cos \phi F_u
\label{eq:beautiful-result}
\end{equation}

Note that the Coriolis effect does not appear explicitly on the right-hand side because it  is absorbed into the material derivative of $m$.\\

This compact equation leads to a simple explanation of the formation of the Jet Stream.  To see this, first note that zonal symmetry allows us to ignore the zonal pressure gradient giving us:

\begin{equation}
\frac{Dm}{Dt} =  a \cos \phi F_u
\label{eq:m-conserve}
\end{equation}  

In the Hadley cell, equatorward air at low levels acquires high angular momentum due to the large planetary radius, while poleward flow aloft conserves this momentum as required by Eq. \ref{eq:m-conserve}. Recall that the specific axial angular momentum is defined as  $m = (u + \Omega_p r \cos \phi) r \cos \phi$, where $\phi$ is latitude. Conservation of $m$ implies that as air moves poleward (decreasing $\cos \phi$), $u$ increases to maintain $m$, forming the jet stream. Friction is the primary non-conservative term.\\

The $m$-equation, $\frac{D m}{Dt} = -\frac{1}{\rho} \frac{\partial p}{\partial \lambda} + a \cos \phi F_u$, serves as a powerful unifying principle for diverse geophysical phenomena. It explains the polar night jet as a consequence of $m$ conservation during poleward flow, links changes in atmospheric angular momentum to variations in the length of day, and attributes the Quasi-Biennial Oscillation to wave-driven torques ($F_u$) in the tropics. In the ocean, it connects poleward flow in the thermohaline circulation to the intensification of currents like the Gulf Stream via $m$ conservation.

\subsection{Meridional Angular Momentum $\ell$}

To explore meridional dynamics we define a meridional angular momentum-like quantity:

\begin{equation}
\ell = av \cos \phi 
\end{equation}

where $ v $ is the meridional velocity. This quantity, while not a conserved angular momentum in the same sense as $ m $, provides insight into the torque balance associated with meridional flows, especially near the equator where $\cos \phi \approx 1$, $\sin \phi \approx 0$, and $\tan \phi \approx 0$. \\

The meridional momentum equation in spherical coordinates for an inviscid, incompressible fluid in the rotating frame (neglecting vertical velocity $ w $ for large-scale flows, as is common in equatorial wave dynamics) is:

\begin{equation}
\frac{D v}{Dt} + 2 \Omega_p \sin \phi \, u + \frac{u^2 \tan \phi}{a} = -\frac{1}{\rho a} \frac{\partial p}{\partial \phi},
\label{eqn:meridional-momentum}
\end{equation}

Here  $ u $ is the zonal velocity, $ p $ is pressure, $\rho$ is density (assumed constant for simplicity), and the terms $ 2 \Omega_p \sin \phi u $ (Coriolis) and $-\frac{u^2 \tan \phi}{a}$ (metric) account for Earth's rotation and spherical geometry. Using Eq. \ref{eqn:meridional-momentum}, the transport equation for $\ell$ can be derived.\\

Multiplying the meridional momentum equation by the lever arm $a \cos \phi$, we obtain the transport equation for the meridional angular momentum, $\ell$:

\begin{equation}
\frac{D \ell}{Dt} = - a \cos \phi f u - u^2 \sin \phi - \frac{\cos \phi}{\rho} \frac{\partial p}{\partial \phi} + a \cos \phi F_v - v^2 \sin \phi
\label{eqn:meridional-ang-momentum}
\end{equation}

where $ f = 2 \Omega_p \sin \phi $ is the coriolis factor. $F_v$ is a friction term which may include viscous and/or eddy viscosity terms.\\

\subsubsection*{Behavior at the Equator}

It may be noted that Eq. \ref{eqn:meridional-ang-momentum} is especially useful for equatorial phenomena where $ \phi \approx 0$ leading to  $\cos \phi \sim 1$, $\sin \phi \sim y/a$, $\tan \phi \sim y/a$, with $\beta = \frac{2 \Omega_p}{a}$ where $y$ is the meridional distance from the equator.  Ignoring terms of $O(y/a)$, we get:

\begin{equation}
\frac{D \ell}{Dt} = -\beta y u - \frac{1}{\rho} \frac{\partial p}{\partial y} + a F_v
\label{eq:ell-equation}
\end{equation}

Further, for some flows we can set $F_v =0$ giving:

\begin{equation}
\frac{D \ell}{Dt} = -\beta y u - \frac{1}{\rho} \frac{\partial p}{\partial y} 
\label{eq:ell-equation-1}
\end{equation}

Eq. \ref{eq:ell-equation-1} gives the balance of $\ell$ transport as balanced by the geostrophic term and the pressure gradient. \\

For some phenomena near the equator (Madden Julien Oscillation (MJO)), the meriodinal velocity, $v=0$ which implies that $\ell =0$.  Therefore Eq. \ref{eq:ell-equation-1} gives a pure geostrophic balance with pressure: 

\begin{equation}
\beta y u - = \frac{1}{\rho} \frac{\partial p}{\partial y} 
\label{eq:ell-equation-2}
\end{equation} 

\subsubsection*{Behavior at the Pole}

Looking at  Eq. \ref{eqn:meridional-ang-momentum} in the limit  $\phi \to 90^\circ$, where $\cos \phi \approx 0$ and  $\sin \phi \approx 1$ gives:

\begin{equation}
\frac{D \ell}{Dt} = -(u^2 + v^2)
\end{equation}

The physical implications of the polar limit become apparent when the right-hand side is re-expressed strictly as a dynamical system in $\ell$. Recalling the definition of meridional angular momentum, $\ell = v a \cos \phi$, we define the local radius of rotation as $R = a \cos \phi$, allowing us to substitute $v = \ell/R$. The polar limit equation transforms into:

\begin{equation}
    \frac{D\ell}{Dt} = -u^2 - \frac{1}{R^2} \ell^2
\end{equation}

This formulation reveals a non-linear, Riccati-type\footnote{Riccati equations which arise in dynamical systems are of the form $y' = q_1(x)y + q_2(x)y^2$.} differential equation dictating the decay of meridional angular momentum. The coefficient of the $\ell^2$ term, $1/R^2$, represents a severe geometric singularity. As a fluid parcel approaches the geographic pole ($R \to 0$), the decay coefficient approaches infinity. \\

Consequently, if a parcel attempts to carry any finite meridional angular momentum ($\ell \neq 0$) to the pole, it experiences infinite deceleration. This dynamical system mathematically enforces the annihilation of cross-polar flow: the fluid is violently sheared, forcing $v \to 0$ and transferring the remaining kinetic energy into the zonal velocity ($u$). This non-linear decay is the exact mathematical mechanism that redirects converging meridional flows into the purely zonal, high-speed atmospheric retaining wall of the polar vortex.\\

This strictly negative definite limit proves that polar dynamics are dominated entirely by the non-linear metric centrifugal force. As poleward-moving air spins up due to angular momentum conservation, the immense outward centrifugal push ($-u^2$) physically prevents the fluid from converging to a singularity at the pole. Instead, the air is forced to pile up into a continuous ring of high-speed zonal winds, creating the impenetrable dynamic wall and the characteristic hollow "eye" of the polar vortex.\\

It may be noted that in spherical coordinates, the Navier-Stokes equations have a coordinate singularity at the poles ($\phi \to 0$), where metric terms such as $\cot \phi$ and $\csc \phi$ diverge to infinity. The  Riccati equation, captures the dynamics at the pole as a limiting case of $\phi \to 0$ and explains the polar vortex collapse.

\subsubsection*{Behavior at Mid-Latitudes}

To isolate the kinematics of the wind vector in the mid-latitudes, we evaluate the meridional momentum balance at exactly $\phi = 45^\circ$.  Setting $\phi =45^\circ$ in Eq.~\ref{eqn:meridional-ang-momentum} gives: 

\begin{equation}
\sqrt{2} \frac{D \ell}{Dt} = - (u^2 + v^2) - a f u - \frac{1}{\rho} \frac{\partial p}{\partial \phi} + a F_v
\end{equation}

Assuming a steady ($D\ell/Dt = 0$) and defining a fixed meridional pressure gradient and friction as the constant $C = \frac{1}{\rho}\frac{\partial p}{\partial \phi} - + a F_v$, the unweighted momentum equation reduces to:

\begin{equation}
u^2 + v^2 + a f u + C = 0
\end{equation}

By completing the square for the zonal velocity $u$, this relationship can be transformed into a strict geometric constraint in the $(u, v)$ velocity space:

\begin{equation}
\left(u + \frac{af}{2}\right)^2 + v^2 = \frac{a^2 f^2}{4} - C
\end{equation}

This mathematically defines a circular curve (a hodograph) centered at $(-af/2, 0)$, whose radius is given by the balance between planetary rotation and the fixed pressure gradient. Any dynamically permissible fluid parcel must possess a velocity vector that originates at the origin and terminates precisely on this circle. Consequently, the total wind velocity $V = \sqrt{u^2 + v^2}$ and the wind direction angle $\theta = \tan^{-1}(v/u)$ are geometrically locked. The atmosphere is forced to select a wind vector that satisfies this circular locus, perfectly illustrating how planetary curvature, the Coriolis parameter, and the pressure gradient universally constrain both the speed and the steering of the mid-latitude flow.\\

To generalize the kinematic constraints of the steady mid-latitude flow, we evaluate the frictionless meridional momentum balance at an arbitrary latitude $\phi \ne 0^\circ, 90^\circ$. Defining the meridional pressure gradient as $C = \frac{1}{\rho}\frac{\partial p}{\partial \phi}$, the balance is given by:

\begin{equation}
- (u^2 + v^2) \sin \phi - a f u \cos \phi - C \cos \phi = 0
\end{equation}

Because the non-linear metric centrifugal accelerations for both the zonal ($u^2$) and meridional ($v^2$) velocities are generated by the isotropic curvature of the sphere, they share identical trigonometric weighting ($\sin \phi$). Dividing the domain by $\sin \phi$ and completing the square for $u$ yields:

\begin{equation}
\left( u + \frac{a f \cot \phi}{2} \right)^2 + v^2 = \frac{a^2 f^2 \cot^2 \phi}{4} - C \cot \phi
\end{equation}

Substituting the planetary Coriolis parameter, $f = 2\Omega_p \sin \phi$, allows a geometric simplification:

\begin{equation}
\left( u + \Omega_p a \cos \phi \right)^2 + v^2 = \Omega_p^2 a^2 \cos^2 \phi - C \cot \phi
\end{equation}

The dynamically permissible velocity locus remains a perfect circle at all latitudes with the origin of this hodograph  centered at $(-\Omega_p a \cos \phi, 0)$ in the $(u, v)$ velocity space. This demonstrates that the geostrophic adjustment of the atmosphere anchors its dynamic wind circle directly against the absolute local rotational velocity of the solid planet. Furthermore, as $\phi \to 0^\circ$, the radius diverges ($\cot \phi \to \infty$), showing the breakdown of gradient wind balance at the equator.\\

The transport of absolute angular momentum ($\ell$) offers a powerful lens for planetary-scale fluid-dynamics. For instance, the anomalous superrotation of the Venusian atmosphere can be physically conceptualized as the continuous, up-gradient spatial pumping of $\ell$ into the equator by planetary waves. Similarly, the termination of the Earth's Hadley cell and the subsequent formation of the subtropical jet emerge naturally as a strict geometric crisis: poleward-moving air must violently accelerate to conserve its rotational debt as the planetary lever arm shrinks. Furthermore, complex, wave-driven phenomena such as the rhythmic reversals of the Quasi-Biennial Oscillation (QBO) can be elegantly recast simply as the alternating vertical deposition of $\ell$ by trapped equatorial waves, demonstrating the utility of the rotational framework across fluid scales.

\section{Compressible Flows}

Building on the incompressible case which has been the subject of this article this far, we now generalize the $\Lvec$ framework to compressible flows. For compressible flows, density variations significantly influence the dynamics, necessitating an appropriate definition of angular momentum to capture the conservation properties effectively. We define the angular momentum per unit volume as:

\begin{equation}
\Lvec = \rvec \times \rho \uvec
\end{equation}

where now $\rho = \rho(\rvec, t)$ is the variable fluid density. The Navier-Stokes equations for a compressible, viscous fluid are:

\begin{equation}
\frac{\partial \rho}{\partial t} + \nabla \cdot (\rho \uvec) = 0,
\label{eqn:continuity-compressible}
\end{equation}

\begin{equation}
\frac{\partial (\rho \uvec)}{\partial t} + (\uvec \cdot \nabla) (\rho \uvec) = -\nabla p + \nabla \cdot \boldsymbol{\tau} - \rho \uvec (\nabla \cdot \uvec)
\label{eqn:momentum-compressible}
\end{equation}

where $ p $ is the pressure, and $\boldsymbol{\tau}$ is the viscous stress tensor. For a Newtonian fluid, $\boldsymbol{\tau} = \mu \left( \nabla \uvec + (\nabla \uvec)^T \right) + \lambda (\nabla \cdot \uvec) \Ivec$, with $\mu$ as the dynamic viscosity, $\lambda$ as the second viscosity coefficient, and $\Ivec$ as the identity tensor.  For the sake of simplicity, we will assume that $\mu$ and $\lambda$ are constants throughout this analysis. The divergence of the stress tensor is:

\[
\nabla \cdot \boldsymbol{\tau} = \mu \nabla^2 \uvec + (\mu + \lambda) \nabla (\nabla \cdot \uvec).
\]

Taking the cross product of Eq. \ref{eqn:momentum-compressible} with $\rvec$, it can be seen that  $ \rvec \times \frac{\partial (\rho \uvec)}{\partial t} + \rvec \times (\uvec \cdot \nabla) (\rho \uvec) =\frac{D \Lvec}{Dt} $.  The angular momentum transport equation can then be written as:

\begin{equation}
\frac{D \Lvec}{Dt} = -\rvec \times \nabla p + \rvec \times (\nabla \cdot \boldsymbol{\tau}) - (\nabla \cdot \uvec) \Lvec
\label{eqn:compressible-angmom}
\end{equation}

The compressibility term $-(\nabla \cdot \uvec) \Lvec$ accounts for density changes affecting angular momentum, a key distinction from incompressible flows.\\

To calculate the moment of the viscous stress in terms of $\mathbf{L}$, we utilize the general identity for the Laplacian of a cross product, $\nabla^2 (\mathbf{A} \times \mathbf{B})$. Expressed in index notation, the product rule distributes the spatial derivatives ($\partial_m \partial_m$) as:
\begin{equation}
    [\nabla^2 (\mathbf{A} \times \mathbf{B})]_k = [(\nabla^2 \mathbf{A}) \times \mathbf{B}]_k + [\mathbf{A} \times (\nabla^2 \mathbf{B})]_k + 2 \epsilon_{kij} [ \partial_m A_i \partial_m B_j ]
\end{equation}
The final term represents the contraction of the respective gradients. Evaluating this identity for the angular momentum vector, we set $\mathbf{A} = \mathbf{r}$ and $\mathbf{B} = \rho \mathbf{u}$. Because the position vector has constant gradients ($\partial_m x_i = \delta_{mi}$) and zero concavity ($\nabla^2 \mathbf{r} = 0$), the first term vanishes and the cross-term elegantly collapses:
\begin{equation}
    2 \epsilon_{kij} [ \delta_{mi} \partial_m (\rho u_j) ] = 2 \epsilon_{kij} \partial_i (\rho u_j) \equiv 2 [\nabla \times (\rho \mathbf{u})]_k
\end{equation}
Substituting this result back into the expanded identity yields:
\begin{equation}
    \nabla^2 \mathbf{L} = \mathbf{r} \times \nabla^2 (\rho \mathbf{u}) + 2 \nabla \times (\rho \mathbf{u})
\end{equation}
Rearranging to isolate the moment of the viscous diffusion term, we obtain the exact relation:
\begin{equation}
    \mathbf{r} \times \nabla^2 (\rho \mathbf{u}) = \nabla^2 \mathbf{L} - 2 \nabla \times (\rho \mathbf{u})
\end{equation}

Using ``dynamic" vorticity (see Truesdell \cite{truesdell2018kinematics}) defined as:

\begin{equation}
\tilde{\omvec} = \nabla \times  \rho \uvec
\end{equation}

We note that the dynamic vorticity does not share the property of being the asymmetric component of the decomposition of the velocity gradient tensor into symmetric and asymmetric parts.  However, it can be interpreted as being the ``spin" angular momentum.  Thus we can write the moment of the shear stress as:

\begin{equation}
\rvec \times  \nabla^2 (\rho \uvec) = \nabla^2 \Lvec - 2 \tilde{\omvec}
\end{equation}

which allows us to write the overall moment of stress as:

\begin{equation}
\rvec \times \nabla \cdot \boldsymbol{\tau} =  \mu(\nabla^2 \Lvec - 2 \tilde{\omvec})  + \rvec \times (\mu + \lambda) \nabla (\nabla \cdot \uvec)
\end{equation}

With this, we can now write the compressible $\Lvec$ equation as:

\begin{equation}
\frac{D \Lvec}{Dt} = -\rvec \times \nabla p + \mu(\nabla^2 \Lvec - 2 \tilde{\omvec})  + \rvec \times (\mu + \lambda) \nabla (\nabla \cdot \uvec) - (\nabla \cdot \uvec) \Lvec
\label{eq:ang-mom-comp-final}
\end{equation}

This is the result we have been seeking.  When we compare Eq. \ref{eq:ang-mom-comp-final} to its incompressible counterpart, Eq. \ref{eqn:angmom-full}, we note that the compressible equation has a similar structure except for:  (i) The ``spin" based diffusion is based on the ``dynamic" vorticity $\tilde{\omvec}$ (ii) There is an extra compression based viscous action term which also includes the bulk modulus $\lambda$  (iii) There is another amplification term related to volumetric expansion/contraction of the type $(\nabla \cdot \uvec) \Lvec$.\\

The arguments presented for the lift generation mechanism in the incompressible flow case, are also applicable here, but we need to work with $\tilde{\omvec}$ and also need to account for the dilatation effects coming from non-zero $\nabla \cdot \uvec$.\\

This compressible $\Lvec$ equation will have interesting applications for curved shock waves like the bow shock which in terrestrial and also astrophysical domains.\\

\subsection{Oblique Shocks as Angular Momentum Line Sources}

\begin{figure}[h!]
    \centering
    \includegraphics[width=0.5\textwidth]{"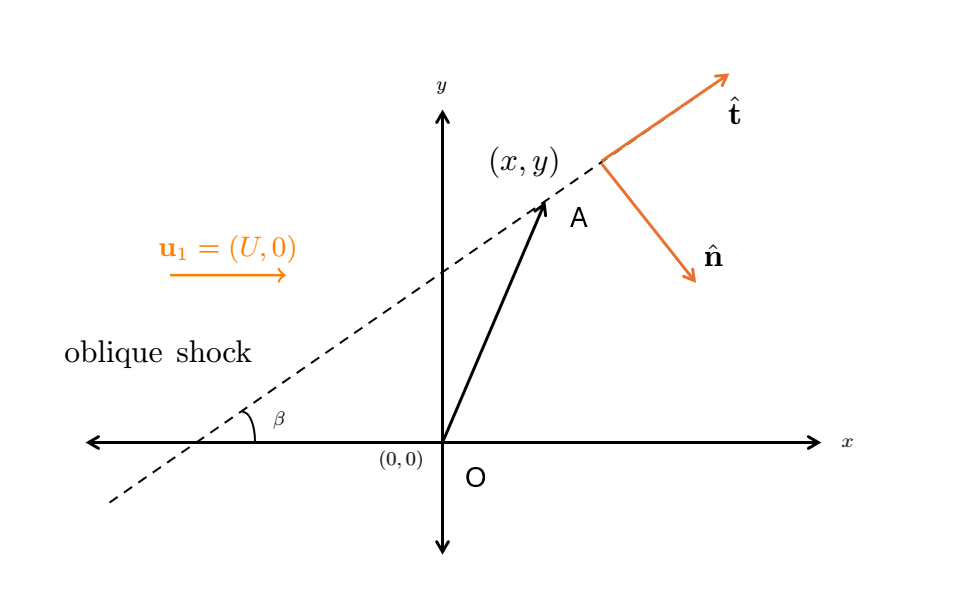"}
    \caption{Oblique shock with a flow impinging at an angle $\beta$.}
    \label{fig:oblique-shock}
\end{figure}

We consider an oblique shock with the geometry given in Figure \ref{fig:oblique-shock} where an oblique shock at an angle $\beta$ to the impinging supersonic with an upstream velocity given by:

\begin{equation}
\uvec_1 =  U \ihat
\end{equation}

The tangential and normal components to the shock are given by the direction vectors $\hat{\nvec} = (\sin \beta, -\cos \beta)$ and $\hat{\mathbf{t}} = (\cos \beta, \sin \beta)$.  The upstream velocity components decomposed along the normal and tangential directions are given by:

\begin{align}
u_{n1} = U \sin \beta \\
u_{t1} = U \cos \beta 
\end{align}

The velocity components downstream of the shock are given by:

\begin{align}
u_{n2} &= \frac{\rho_1}{\rho_2} u_{n1}= \frac{\rho_1}{\rho_2} U \sin \beta \\
u_{t2} &= u_{t1}
\end{align}

where $\frac{\rho_1}{\rho_2}$ can be obtained from the Rankine-Hugoniot relations. This now allows us to write the downstream velocity, $\uvec_2$ as:

\begin{equation}
\uvec_2 =  U \left( \sin^2 \beta \frac{\rho_1}{\rho_2} + \cos^2 \beta \right) \ihat + U \sin \beta \cos \beta \left( 1 - \frac{\rho_1}{\rho_2}  \right) \jhat
\end{equation}

The angular momentum per unit volume is $ L_z = \mathbf{r} \times (\rho \uvec) $, so we can define the $L_{z1}$ and $L_{z2}$ upstream and downstream angular momenta as:

\begin{align}
L_{z1} &= \rho_1 (x v_1 - y u_1) = -\rho_1 y U \\
L_{z2} &= \rho_2 (x v_2 - y u_2)
\end{align}

The jump is given by:
$$  [L_z] = L_{z2} - L_{z1} = \rho_2 (x v_2 - y u_2) + \rho_1 y U $$

After substituting for $u_2,v_2$ and simplification, we obtain:

\begin{equation}
[L_z] = \rho_2\left( \frac{\rho_1}{\rho_2} - 1 \right) u_{n1} (x \cos \beta - y \sin \beta).
\label{eq:oblique-shock-1}
\end{equation}

Thus, the oblique shock acts as a singular source of angular momentum.  From normal momentum conservation:

\begin{equation}
[p] = \rho_1 u_{n1}^2 \left(1 - \frac{\rho_1}{\rho_2}\right),
\end{equation}

so we can define:

$$  \frac{\rho_1}{\rho_2} - 1 = -\frac{[p]}{\rho_1 u_{n1}^2} $$

Substitute into Eq. \ref{eq:oblique-shock-1} to obtain:

\begin{equation}
[L_z] = \rho_2\frac{[p]}{\rho_1 u_{n1}} (y \sin \beta - x \cos \beta).
\label{eq:oblique-shock-2}
\end{equation}  

This confirms the shock as a delta source of $ L_z $, with magnitude tied to pressure jump, normal velocity, and geometry.

\section{Stokes Flows \& Rotlets}

In Stokes flows, characterized by low Reynolds numbers ($\text{Re} \ll 1$),  viscous forces dominate over inertial effects, and the latter terms are ignored.  For general references on Stokes flows, Leal \cite{leal2007advanced}, Currie \cite{currie2002fundamental} may be consulted.\\

The governing equation for stokes flow is given by:

\begin{equation}
\nabla p = \mu \nabla^2 \uvec, \quad \nabla \cdot \uvec = 0,
\label{eq:stokes-equations}
\end{equation}

The angular momentum transport equation for $\Lvec$ is derived by taking the cross product of the Stokes momentum equation above (Eq. \ref{eq:stokes-equations}), we obtain:

\begin{equation}
 \rvec \times \nabla p = \mu \left( \nabla^2 \Lvec - 2 \omvec \right).
\label{eq:L-stokes-steady}
\end{equation}

This equation represents the balance between the pressure torque $-\rvec \times \nabla p$ and the viscous torque $\mu (\nabla^2 \Lvec - 2 \omvec)$. The viscous torque decomposes into a diffusive component ($\mu \nabla^2 \Lvec$) that redistributes angular momentum spatially and a local spin torque ($-2 \mu \omvec$) that dissipates angular momentum proportional to the vorticity. This decomposition, unique to the $\Lvec$ framework, is particularly insightful for Stokes flows, where rotational effects are critical in microscale dynamics.\\

It is well known that Stokeslets correspond to Greens functions of Eq. \ref{eq:stokes-equations}  where $\nabla p  = \delta(\rvec)$. We will show that Rotlets (see Chwang et al. \cite{Chwang1975}), which have been well known correspond to being the Greens functions for Eq. \ref{eq:L-stokes-steady}.  \\

Rotlets are a fundamental solutions to the Stokes equations corresponding to a point torque applied at a point in the fluid. Rotlets model the flow field generated by localized rotational disturbances, such as a rotating particle, a microorganism's flagellum, or a ciliary beat. To derive the rotlet velocity field, consider a point torque $\mathbf{T}$ applied at the origin. The Stokes equations are linear, allowing the use of Green's functions or the Oseen tensor to solve for the velocity field induced by a singular torque.\\

The velocity field for a rotlet, first defined by Chwang et al. \cite{Chwang1975} with torque $\mathbf{T}$ is given by:

\begin{equation}
\uvec_{\text{rotlet}}(\rvec) = \frac{\mathbf{T} \times \rvec}{8 \pi \mu r^3},
\label{eq:rotlet-velocity}
\end{equation}

where $r = |\rvec|$ is the distance from the origin.  This solution satisfies the Stokes equations away from the origin, as can be verified by checking incompressibility and the momentum balance:\\

Incompressibility: $\nabla \cdot \uvec_{\text{rotlet}} = \nabla \cdot \left( \frac{\mathbf{T} \times \rvec}{8 \pi \mu r^3} \right) = 0$, since the divergence of a curl is zero.\\
Momentum balance: Compute $\nabla^2 \uvec_{\text{rotlet}}$ and ensure it matches a pressure gradient plus a singular term at the origin.\\

The vorticity of the rotlet is computed as:

\begin{equation}
\omvec_{\text{rotlet}} = \nabla \times \uvec_{\text{rotlet}}.
\end{equation}

Using the vector identity for the curl of a cross product, $\nabla \times (\mathbf{a} \times \mathbf{b}) = \mathbf{a} (\nabla \cdot \mathbf{b}) - \mathbf{b} (\nabla \cdot \mathbf{a}) + (\mathbf{b} \cdot \nabla) \mathbf{a} - (\mathbf{a} \cdot \nabla) \mathbf{b}$, and noting that $\mathbf{T}$ is constant and $\rvec/r^3$ involves a radial gradient, and using $\nabla \cdot \mathbf{T} = 0$, $(\rvec \cdot \nabla) \mathbf{T} = 0$, $\nabla \cdot \rvec = 3$, and $\nabla \frac{1}{r^3} = -\frac{3 \rvec}{r^5}$, we obtain:

\begin{equation}
\omvec_{\text{rotlet}} = \frac{3 (\mathbf{T} \cdot \rvec) \rvec - r^2 \mathbf{T}}{8 \pi \mu r^5}.
\label{eq:rotlet-vorticity}
\end{equation}

This vorticity field decays as $1/r^4$, indicating a concentrated rotational effect near the origin. The corresponding angular momentum field for the rotlet is:

\begin{equation}
\Lvec_{\text{rotlet}} = \rvec \times \uvec_{\text{rotlet}} = \rvec \times \left( \frac{\mathbf{T} \times \rvec}{8 \pi \mu r^3} \right).
\end{equation}

Using the vector triple product identity $\mathbf{a} \times (\mathbf{b} \times \mathbf{c}) = \mathbf{b} (\mathbf{a} \cdot \mathbf{c}) - \mathbf{c} (\mathbf{a} \cdot \mathbf{b})$, we compute: $\rvec \times (\mathbf{T} \times \rvec) =  r^2 \mathbf{T} - (\mathbf{T} \cdot \rvec) \rvec$.  Thus giving:

\begin{equation}
\Lvec_{\text{rotlet}} = \frac{r^2 \mathbf{T} - (\mathbf{T} \cdot \rvec) \rvec}{8 \pi \mu r^3}.
\label{eq:L-rotlet}
\end{equation}

This $\Lvec_{\text{rotlet}}$ field decays as $1/r$, slower than the Stokeslet velocity ($1/r$) or its vorticity ($1/r^2$), reflecting the long-range influence of rotational disturbances in viscous flows. The slower decay implies that torques propagate farther than forces, which is critical for understanding interactions in dense suspensions or microbial systems where rotational effects dominate over linear forces.\\

To verify the role of the rotlet in the angular momentum transport equation, we evaluate the viscous torque term in Eq. \ref{eq:L-stokes-steady}:

\[
\mu (\nabla^2 \Lvec_{\text{rotlet}} - 2 \omvec_{\text{rotlet}}).
\]

Away from the origin, the rotlet velocity is harmonic ($\nabla^2 \uvec_{\text{rotlet}} = 0$, since $\nabla^2 (\mathbf{T} \times \rvec / r^3) \propto \nabla^2 (1/r) \propto \delta(\rvec)$), and thus $\nabla^2 \Lvec_{\text{rotlet}} = \rvec \times \nabla^2 \uvec_{\text{rotlet}} + 2 \omvec_{\text{rotlet}} = 2 \omvec_{\text{rotlet}}$. Therefore:

\[
\nabla^2 \Lvec_{\text{rotlet}} - 2 \omvec_{\text{rotlet}} = 0, \quad r \neq 0.
\]

At the origin, the singularity in $\uvec_{\text{rotlet}}$ introduces a delta function. The Stokes momentum equation for the rotlet gives $\nabla p = \mu \nabla^2 \uvec_{\text{rotlet}} = -\mathbf{T} \delta(\rvec)$, and crossing with $\rvec$:

\begin{equation}
\mu (\nabla^2 \Lvec_{\text{rotlet}} - 2 \omvec_{\text{rotlet}}) = \mathbf{T} \delta(\rvec).
\end{equation}

This confirms that the rotlet acts as the Green's function for the torque balance in Eq. \ref{eq:L-stokes-steady}, with a singular torque source at the origin. This property makes the $\Lvec$ framework ideal for analyzing rotational flows, as it directly captures the torque-driven dynamics without needing to solve for vorticity or velocity gradients separately.\\

The $\Lvec$ framework has potential applications in other Stokes flows and microfluidics, where viscous torques dominate at low Reynolds numbers.

\section{Conclusions}

We have presented a comprehensive transport framework for the angular momentum density field $\Lvec$, demonstrating its utility as a complementary lens to the classical velocity-vorticity formulation. By taking the moment of the Navier-Stokes equations, we unveiled a natural decomposition of the viscous torque into a spatial diffusion term for $\Lvec$ and a local dissipative drag proportional to vorticity. This separation reveals that vorticity acts as the fundamental kinematic source for the $\Lvec$ field, providing a deterministic explanation for lift as the reaction to the turning of the flow—a mechanism that bypasses the abstract requirements of the Kutta condition in viscous regimes.\\

A primary achievement of this work is the reformulation of the hydrodynamic impulse. The resulting triple decomposition yields a clean separation of terms into dilatational, volumetric, and rotational flux components. Unlike traditional methods that rely on irrotational assumptions, this formulation permits the direct calculation of the viscous added mass force, $\Fvec_{\text{nc}}$, which accounts for the inertial resistance of the fluid while naturally incorporating the effects of boundary layers and separated wakes.\\

In the domain of geophysical flows, the $\Lvec$ framework offers a significant conceptual simplification. While traditional vorticity-based models treat planetary rotation as a ``virtual'' or background vorticity ($f$) that must be artificially reconciled with relative motion, the axial angular momentum $m$-equation absorbs these Coriolis effects natively. This avoids the need for potential vorticity decompositions, framing planetary-scale phenomena—from jet streams to western boundary currents—as simple, explicit torque balances on a rotating sphere.\\

Furthermore, we have shown that the framework remains robust across diverse flow regimes. In Stokes flows, the rotlet is established as the Green’s function for the $\Lvec$ transport equation, while in compressible flows, oblique shocks are revealed as singular line sources of angular momentum tied to the pressure jump. Ultimately, the $\Lvec$ formalism provides the necessary kinematic closure to unify added mass and circulatory lift within a single, dimensionally consistent impulse budget, offering a unified paradigm for modern fluid dynamics research.\\

This new framework opens new avenues for future research in domains that have not been touched upon in this article such as turbulence.

\bibliographystyle{siam}
\bibliography{f-bib}

\end{document}